\documentclass[lettersize,journal]{IEEEtran}
\usepackage{amsmath,amssymb,amsfonts}
\usepackage{algorithmic}
\usepackage{algorithm}
\usepackage{array}
\usepackage[caption=false,font=footnotesize,labelfont=rm,textfont=rm]{subfig}
\usepackage{stfloats}
\usepackage{url}
\usepackage{verbatim}
\usepackage{graphicx}
\usepackage{cite}
\usepackage[pagewise]{lineno}
\usepackage{booktabs}
\usepackage{bm}
\usepackage{enumitem}
\usepackage{multicol}
\usepackage{multirow}
\usepackage{tabularray}
\usepackage{tabularx}
\usepackage{threeparttable}
\usepackage{makecell}
\usepackage{placeins}
\usepackage{orcidlink}
\usepackage{siunitx}
\usepackage{overpic}
\usepackage{soul}
\usepackage{xcolor}
\usepackage{preprint}

\DeclareSIUnit{\sample}{Sa}
\DeclareSIUnit{\gauss}{Gs}

\begin{document}


\title{%
Realization of Precise Perforating Using Dynamic Threshold and \linebreak
Physical Plausibility Algorithm for Self-Locating Perforating \linebreak
in Oil and Gas Wells
}

\author{%
Si-Yu~Xiao      \orcidlink{0009-0006-3095-3242},
Guo-Hui~Ren     \orcidlink{0009-0003-8514-095X},
Tian-Hao~Mao    \orcidlink{0009-006-1898-1080},
Yu-Qiao~Chen    \orcidlink{0009-0000-7241-8850},
Yi-An~Liu       \orcidlink{0000-0003-0123-3343},
Jun-Jie~Wang    \orcidlink{0000-0001-7183-422X},\linebreak
Kai~Tang        \orcidlink{0009-0006-4764-3435},
Xin-Di~Zhao     \orcidlink{0009-0007-4304-001X},
Zhi-Jian~Yu     \orcidlink{0009-0001-5099-482X},
Shuang~Liu      \orcidlink{0000-0002-0587-4415}\IEEEauthorrefmark{1},
Tu-Pei~Chen     \orcidlink{0000-0002-1098-9575} and
Yang~Liu        \orcidlink{0000-0003-0615-7036}

\thanks{S.Y.~Xiao, T.H.~Mao, Y.Q.~Chen, Y.A.~Liu, J.J.~Wang, S.~Liu and Y.~Liu are with the State Key Laboratory of Electronic Thin Films and Integrated Devices, University of Electronic Science and Technology of China, Chengdu 611731, China.}
\thanks{G.H.~Ren, K.~Tang, X.D.~Zhao are with the Southwest Branch of China National Petroleum Corporation Logging Co., Ltd., Chongqing 401100, China.}
\thanks{Z.J.~Yu is with Chengdu Original Dynamics Technology Co. Ltd., Chengdu 610041, China.}
\thanks{T.P.~Chen is with the School of Electrical and Electronic Engineering, Nanyang Technological University, Singapore 639798.}
\thanks{This work is supported by NSFC under project No.~62404033 and 62404034. This work is also supported by China National Petroleum Corporation Logging Co., Ltd. (CNLC) under project No.~CNLC2023-7A01.}
\thanks{\IEEEauthorrefmark{1}Corresponding author.}
}



\maketitle

\copyrightnotice

\begin{abstract}
Accurate depth measurement is critical for targeting designated perforation intervals to maximize hydrocarbon recovery. While next-generation automated wireless perforating techniques reduce reliance on costly surface infrastructure and personnel, they lack the continuous depth correlation provided by conventional wireline cables. Consequently, correlating real-time casing collar locator (CCL) signals with a pre-recorded casing tally is essential for automatic depth determination. However, implementing this measurement remains challenging: downhole instruments must process CCL signals in real-time to identify collar signatures from complex interference, a task severely restricted by the limited computational resources and power budget of high-temperature downhole electronics.
To address these constraints, this work proposes the Dynamic Threshold and Physical Plausibility Depth Measurement and Perforation Control (DTPPMP) system. This integrated solution enables in situ depth calibration by correlating CCL signals with the casing tally using lightweight algorithms for dynamic-threshold-based collar recognition and physical plausibility verification.
Field tests demonstrate a collar recognition F1-score of \SI{98.6}{\percent} at a throughput of \SI{1000}{\sample\per\second}. Notably, the algorithm requires only \SI{1.5}{\micro\second} per sample, confirming its computational efficiency and suitability for deployment on resource-constrained, high-temperature downhole platforms.
\end{abstract}

\begin{IEEEkeywords}
Automatic Perforation, Automatic System, Casing collar locator, Depth Measurement, Measurement System, Self-Locating Perforating
\end{IEEEkeywords}

\FloatBarrier

\section{Introduction}\label{sec:intro}

\begin{figure}[!tb]
    \centering
    \subfloat[]{\includegraphics[width=0.7\columnwidth]{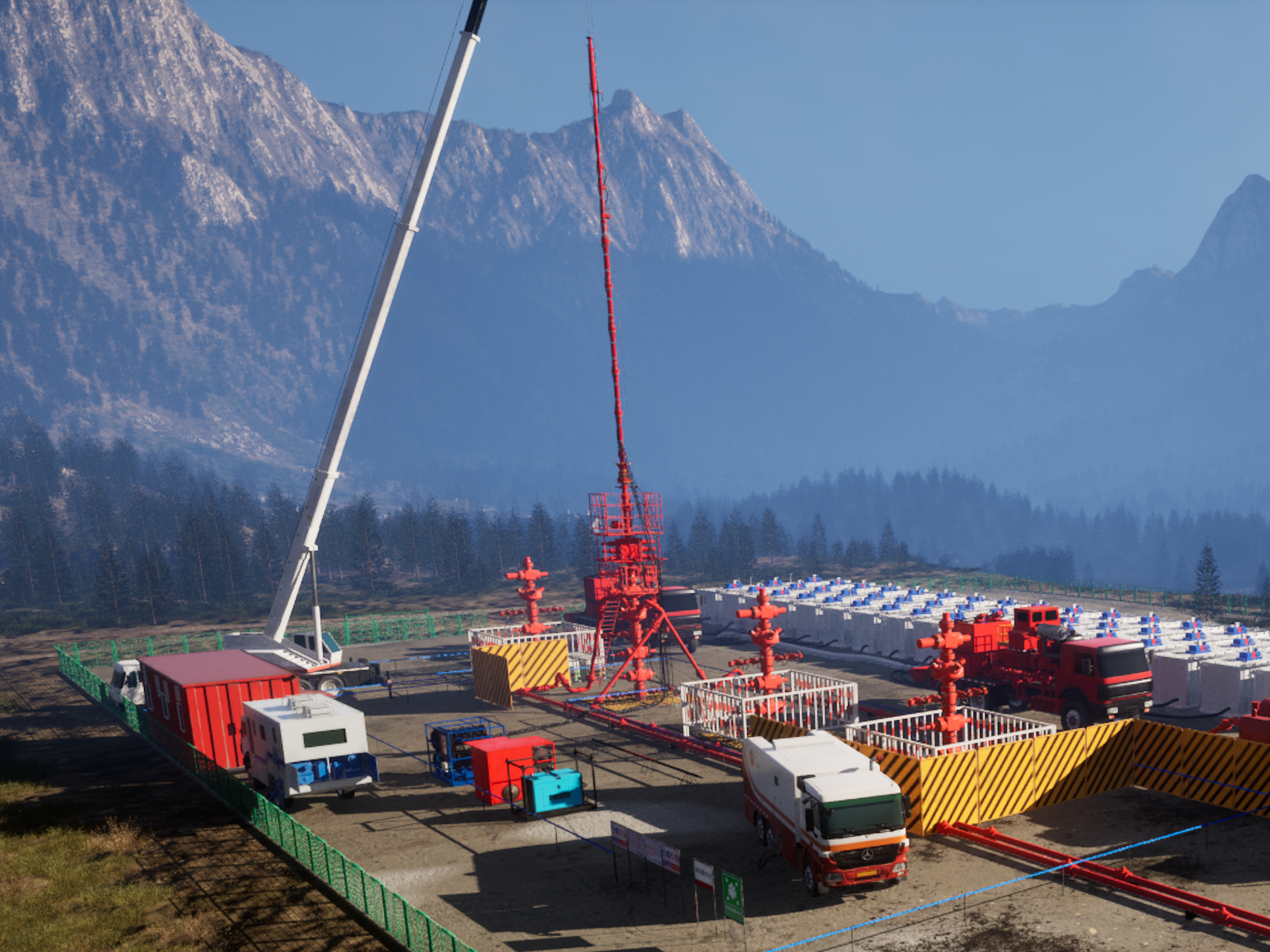}\label{fig1a}}
    \linebreak
    \subfloat[]{\includegraphics[width=0.4\columnwidth]{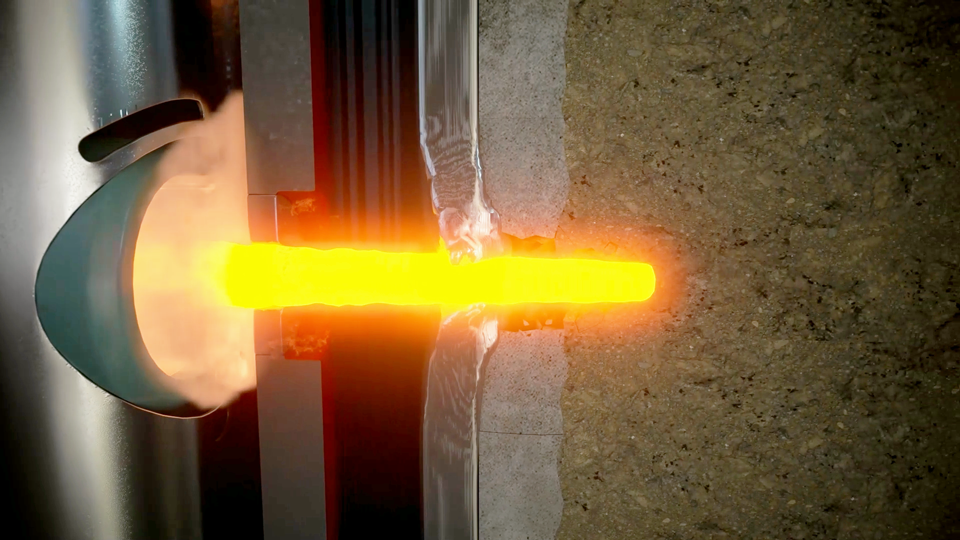}\label{fig1b}}
    \hspace{0.05\columnwidth} 
    \subfloat[]{\includegraphics[width=0.4\columnwidth]{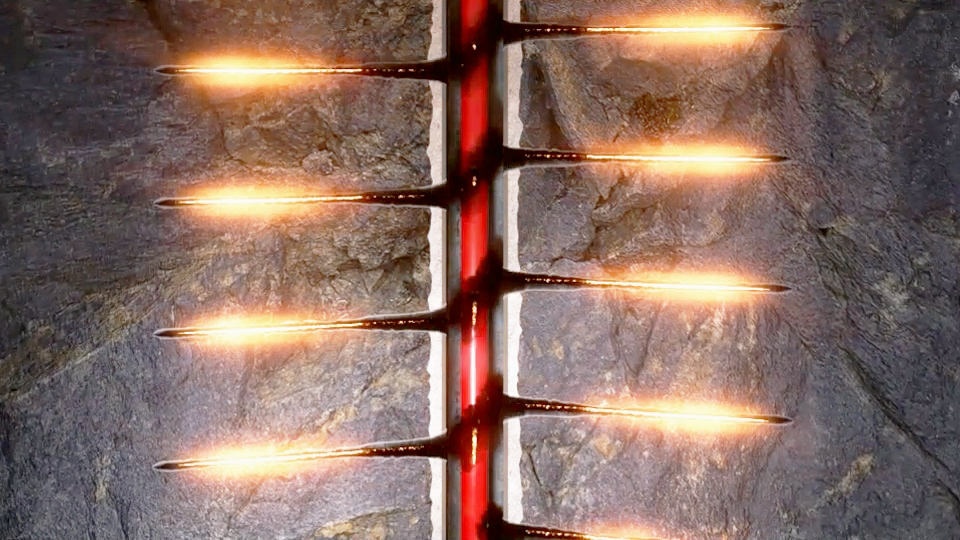}\label{fig1c}}
    \linebreak
    \subfloat[]{\includegraphics[width=0.4\columnwidth]{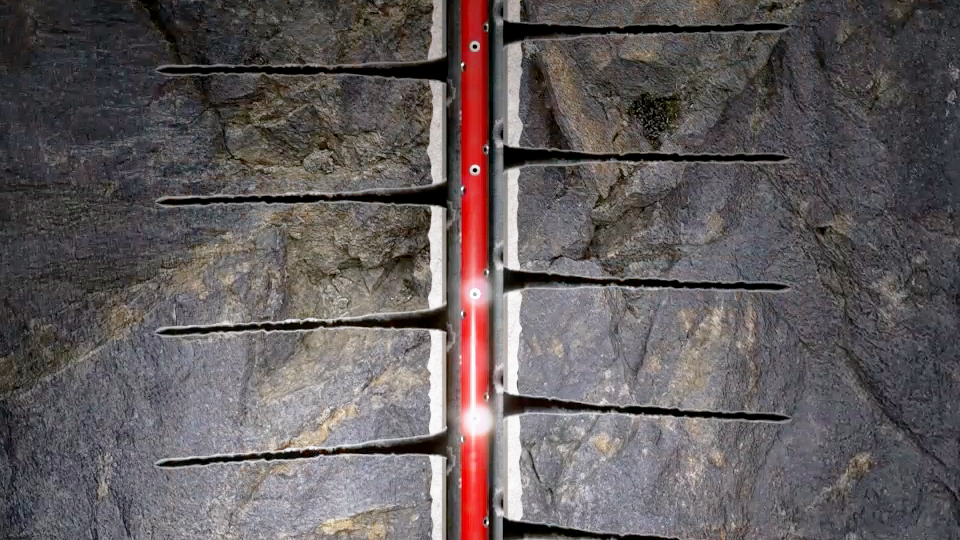}\label{fig1d}}
    \hspace{0.05\columnwidth}
    \subfloat[]{\includegraphics[width=0.4\columnwidth]{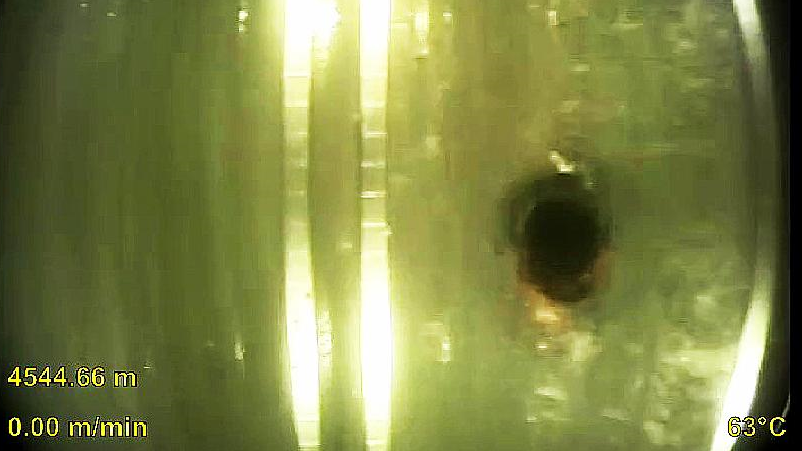}\label{fig1e}}
    \caption{
        \textbf{(a)}~Surface view of the development process for a typical oil and gas well;
        \textbf{(b)}~Detonation of a shaped charge;
        \textbf{(c)}~Perforation by a perforating gun;
        \textbf{(d)}~Tunnels created by a perforating gun in rock formations;
        \textbf{(e)}~Entrance hole on the casing captured by a downhole camera.
        Images courtesy of CNLC.
    }
    \label{fig1}
\end{figure}

\IEEEPARstart{P}{erforation} is a pivotal operation in the development of oil and gas resources. It establishes a connection between the wellbore and the reservoir \cite{liu2014oilwell} by creating tunnels in the rock formations employing shaped charges \cite{lu2012oilandgasfield}, as shown in Fig.~\ref{fig1}. Perforating at optimal intervals maximizes hydrocarbon recovery from subsurface reservoirs, thereby directly influencing well productivity \cite{harris1966effect,lu2012oilandgasfield,liu2014oilwell,deffenbaugh2017untethered}. Consequently, the accuracy of perforation positioning is critical \cite{liu2014oilwell,raman2024data}.
However, accurate downhole navigation (specifically, measuring the depth of downhole equipment) presents significant challenges due to the harsh operating environment. This environment is characterized by wellbores with restricted diameters (often only a few feet) that extend thousands of meters underground \cite{zhao2021highspeed,seren2022miniaturized}, coupled with high-temperature and high-pressure (HPHT) conditions \cite{ambrus2005digital,mijarez2014hpht}.

~

\begin{figure}[!htbp]
    \centering
    \includegraphics[width=0.75\columnwidth]{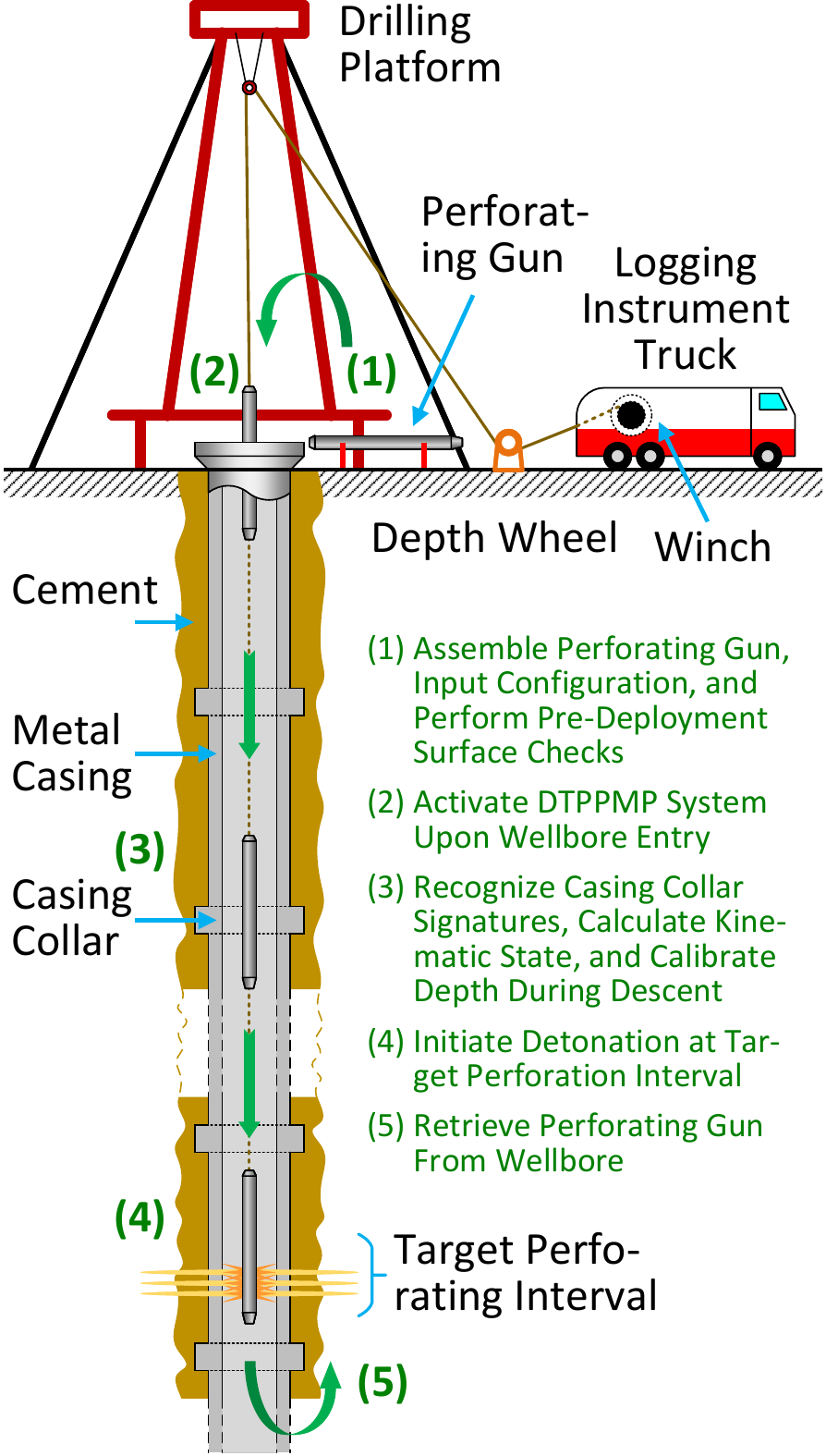}
    \caption{
        Cross-sectional illustration of a typical oil and gas well structure and the operational workflow of the DTPPMP system.
    }
    \label{fig2}
\end{figure}

Tethered conveyance constitutes a conventional technique in downhole operations, utilizing a physical cable to connect downhole instrumentation to the surface \cite{seren2022miniaturized}. This cable, typically composed of armored electromechanical wire, facilitates power transmission, data telemetry, and mechanical support \cite{seren2022miniaturized}. Among these applications, wireline perforating stands out as a predominant approach in current production practices \cite{seren2023magnetic}.
Fig.~\ref{fig2} depicts a typical hydrocarbon well site. During wireline perforating, a toolstring comprising various sensors and perforating guns is lowered into the wellbore via a winch located on the drilling platform. The tool depth is initially estimated by surface wheel measurement (SWM) method. Sensor signals are transmitted in real-time to the surface logging system through the cable \cite{mijarez2014hpht,alvarez2018theory}, allowing for depth correlation and correction. Upon reaching the target interval, the shaped charges within the gun are initiated to execute the perforation.

Wireline perforating is a well-established technique offering distinct advantages, including:
\begin{enumerate}
\item Robustness in high-temperature and high-pressure (HPHT) environments \cite{mijarez2014hpht};
\item Precise control over the running in hole (RIH) speed \cite{seren2022miniaturized};
\end{enumerate}
Conversely, the technique encounters several challenges:
\begin{enumerate}
\item Operational complexity due to the physical cable, necessitating heavy infrastructure such as logging trucks, lifting equipment, and specialized crews \cite{entchev2011autonomous,seren2018untethered,seren2019wireless};
\item Elevated safety risks associated with the complexity of equipment and operations \cite{entchev2011autonomous,angeles2012justintime};
\item Increased operational costs driven by logistical requirements and equipment maintenance \cite{entchev2011autonomous,deffenbaugh2017untethered}.
\end{enumerate}

To address these limitations, untethered conveyance has emerged as a promising alternative. Extensive research has focused on untethered logging instrumentation \cite{deffenbaugh2017untethered,seren2018untethered,seren2019wireless,seren2022miniaturized,larbizeghlache2022sensorball} and downhole robotics \cite{seren2023magnetic}. In the context of perforation, ExxonMobil have proposed untethered, self-navigating, and self-destructing disposable perforating tools \cite{entchev2011autonomous} to replace conventional wireline systems.

Wireless perforating offers compelling operational advantages:
\begin{enumerate}
\item Elimination of surface infrastructure and personnel \cite{entchev2011autonomous,seren2019wireless,seren2025probabilistic}, thereby reducing operational costs \cite{entchev2011autonomous,seren2022miniaturized} and safety risks \cite{entchev2011autonomous,angeles2012justintime};
\item Simplified deployment, potentially requiring only a single operator without specialized crews \cite{seren2019wireless};
\item Reduction in environmental footprint and air emissions due to the removal of heavy surface equipment \cite{entchev2011autonomous}.
\end{enumerate}
However, the implementation of this technology is impeded by technical challenges:
\begin{enumerate}
\item Navigational difficulties within the wellbore due to the absence of continuous depth correlation provided by the wireline \cite{seren2022miniaturized,seren2023magnetic};
\item Challenges in real-time wireless data transmission caused by severe attenuation of electromagnetic and acoustic signals over long distances \cite{seren2025probabilistic};
\item Operational constraints in procedures such as pump-down perforating (e.g., Plug-and-Perf), which demand real-time localization and decision-making \cite{ahmed2021case};
\item The unavailability of SWM-based depth data, which necessitates a shift from depth-driven logging to time-driven logging paradigms.
\end{enumerate}
Addressing these challenges is a prerequisite for establishing wireless perforating as a viable industrial solution.

~

Depth determination, also referred to as localization, is a fundamental prerequisite for effective navigation. Several methodologies have been established to address this challenge.

The SWM technique is a direct and straightforward approach for estimating the depth of downhole equipment \cite{alvarez2018theory,raman2024data}. This method relies on quantifying the deployed cable length via a depth wheel (also known as a counter wheel) \cite{alvarez2017design}. However, the accuracy of this approach is susceptible to errors introduced by cable slippage, elastic stretch, calibration discrepancies of the depth wheel, and thermal or viscoelastic deformations affecting both the perforating guns and the cable \cite{raman2024data}.

Natural gamma ray (GR) detection serves as another standard method for depth definition based on formation lithology \cite{seren2022miniaturized,seren2023magnetic,raman2024data}. Nevertheless, the application of GR tools is constrained by the requirement for reduced logging speeds to ensure sufficient signal integration \cite{lu2012oilandgasfield}. Furthermore, high-precision GR instrumentation entails significant capital expenditure, and the retrieval of tools further escalates operational costs.

Inertial navigation utilizing inertial measurement units (IMUs) constitutes a third widely adopted technique \cite{li2015infield,park2018resilient}. However, the IMUs are susceptible to degradation from mechanical shock, vibration, and sensor noise. Moreover, the inherent accumulation of drift errors over extended distances presents a critical challenge that is often uncorrectable without external reference \cite{park2018resilient}.

A wellbore casing string is comprised of individual metal pipe segments connected by casing collars. The cumulative depth of each collar, collectively known as the ``casing tally'', is derived from well completion records, which document the sequential arrangement of pipe installed \cite{seren2025probabilistic}. As the casing structure remains static following installation, the casing tally serves as a robust fiducial reference for downhole depth measurement and navigation \cite{li2013casing,alvarez2018theory,seren2022miniaturized,gidado2023well}.
The casing collar locator (CCL), a sensor frequently integrated into downhole toolstrings \cite{seren2025probabilistic}, is typically constructed with two strong permanent magnets and an induction coil \cite{mijarez2014hpht,alvarez2017design,seren2022miniaturized}. The CCL operates by detecting variations in ferromagnetic mass at the casing collars. As the tool traverses a collar \cite{li2013casing,seren2022miniaturized}, the resulting perturbation in magnetic flux induces an electromotive force (EMF) within the coil \cite{li2013casing,mijarez2014hpht,alvarez2017design,seren2023magnetic,raman2024data}. Consequently, the accurate identification and correlation of these collar signatures is critical for precise depth correction \cite{raman2024data}.
In wireline-conveyed operations, the CCL signal is transmitted in real-time to the surface logging system via the wireline cable \cite{alvarez2018theory}. Conversely, in wireless or autonomous applications, the signal must be processed in situ by the downhole equipment.

Navigation based on CCL signals is a robust and cost-effective methodology \cite{alvarez2017design,alvarez2018theory} that has served as a primary standard in the hydrocarbon industry for decades \cite{seren2022miniaturized}. Additionally, CCL sensors are capable of reliable operation under HPHT conditions \cite{mijarez2014hpht}. Furthermore, the detection mechanism remains effective across varying RIH speeds, a critical feature for untethered equipment where velocity control may fluctuate \cite{seren2022miniaturized}.
However, the traditional workflow, which relies on manual visual correlation by experts, is labor-intensive, subjective, and time-consuming \cite{raman2024data}. This manual approach often results in reduced operational efficiency and increased susceptibility to human error, primarily because raw CCL signals exhibit significant complexity and variability, necessitating analysis to discriminate genuine collar signatures from interference.

\begin{figure}[!t]
    \centering
    \includegraphics[width=\linewidth]{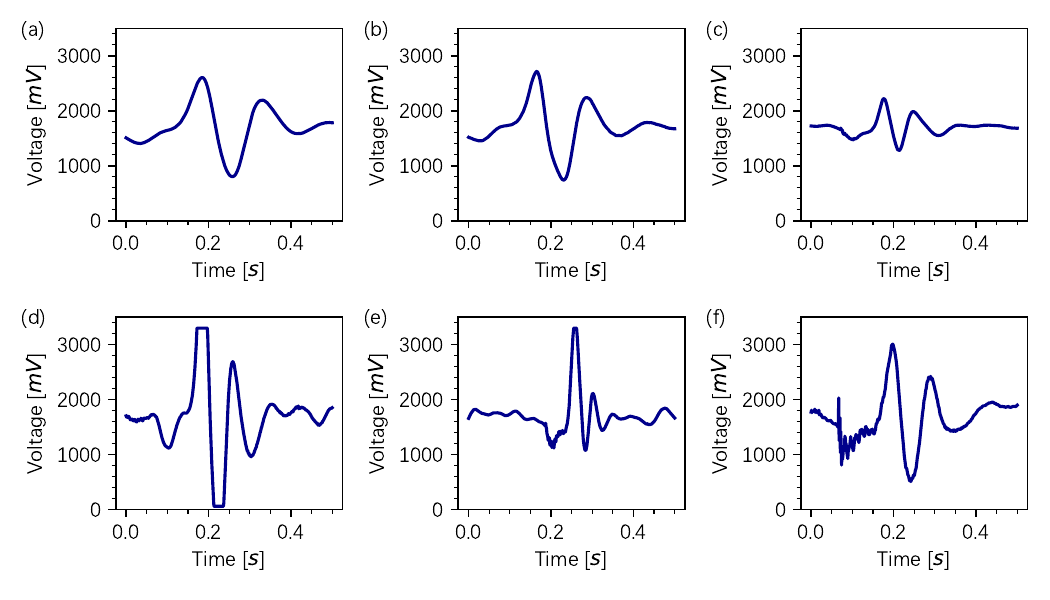}
    \caption{
        Typical casing collar signals:
        \textbf{(a)--(c)}~Clear signals;
        \textbf{(d)}~Signal with excessively large amplitude;
        \textbf{(e)--(f)}~Signals with interference.
    }
    \label{fig3}
\end{figure}

\begin{figure}[!t]
    \centering
    \includegraphics[width=1\linewidth]{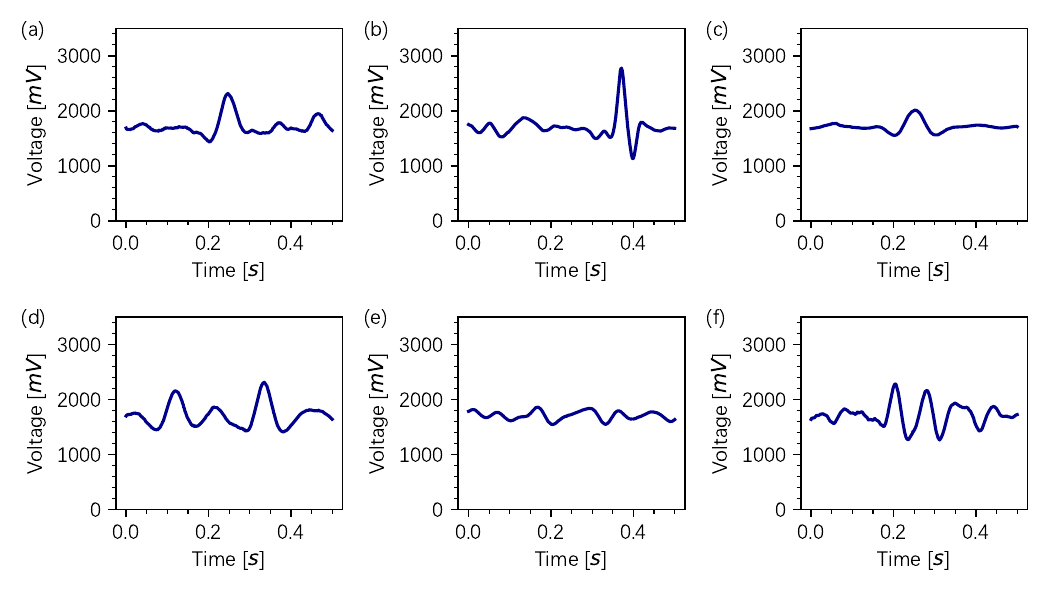}
    \caption{
        Typical interference signals:
        \textbf{(a)--(c)}~Interference with a single large spike;
        \textbf{(d)--(f)}~Interference with continuous spikes.
    }
    \label{fig4}
\end{figure}

Theoretically, casing collars generate characteristic M-shaped bipolar waveforms, hereafter referred to as ``casing (collar) signals'' or ``collar (magnetic) signatures''. Representative examples of genuine collar signatures and interference signals are presented in Figs.~\ref{fig3} and~\ref{fig4}, respectively. In practical scenarios, signal interference arises from diverse sources, which can be categorized as follows:
\begin{enumerate}
\item \textit{Casing Condition and Magnetization:} Mechanical wear, geometric deformation induced by formation pressure, and residual magnetization of the casing string. These factors alter the local magnetic permeability, introducing anomalous signals \cite{alvarez2018theory,zeng2022cclsignal};
\item \textit{System Properties:} Intrinsic characteristics of the downhole equipment (e.g., the CCL sensor) and the physical attributes of the casing (e.g., diameter, wall thickness, and material) \cite{raman2024data};
\item \textit{Transmission Limitations:} Signal attenuation inherent to extended wireline cables \cite{brown1990effects} and degradation due to parasitic circuit parameters, which is particularly detrimental in unmodulated analog transmission systems;
\item \textit{Toolstring Dynamics:} Motion artifacts resulting from the toolstring swinging, rotating, or impacting the casing wall during conveyance;
\item \textit{Operational Factors:} Human errors such as inappropriate amplifier gain settings leading to signal saturation, or non-standard operating procedures \cite{alvarez2018theory,zeng2022cclsignal};
\item \textit{Environmental Noise:} Electromagnetic interference (EMI) and electronic disturbances, including antenna effects coupled into long wireline cables;
\item \textit{Measurement Errors:} Stochastic noise and quantization errors \cite{torrescaceres2022automated}.
\end{enumerate}
Collectively, these factors compromise the detection accuracy of casing collars. Detection errors can lead to erroneous depth corrections, resulting in severe operational consequences \cite{raman2024data}.

It is important to distinguish this interference from stochastic noise, although the term ``noise'' is occasionally used colloquially in the literature to describe these artifacts \cite{seren2025probabilistic}. Unlike stochastic noise, the physical generation mechanism of certain interference signals is analogous to that of genuine collar signatures, making them difficult to discriminate, as evidenced by the similarities observed in Figs.~\ref{fig3} and~\ref{fig4}. Consequently, conventional filtering techniques are often ineffective in suppressing such interference.

~

To address the challenge of accurate casing collar identification, various methodologies have been proposed in recent years. These can be broadly categorized into time-frequency analysis, deterministic time-domain filtering, data-driven approaches, and image-based feature extraction.

Time-frequency analysis represents a conventional approach, utilizing techniques such as the Fourier transform \cite{li2020application} and wavelet transform \cite{li2010approach,li2013casing}. However, these methods often exhibit limited robustness against interference in complex industrial environments \cite{zeng2022cclsignal}. Furthermore, increasing scale resolution to enhance analysis accuracy imposes a significant computational burden, often exceeding the processing capabilities of standard downhole instrumentation.

Deterministic time-domain filtering constitutes another primary methodology. Threshold-based techniques, employing either fixed \cite{wang2006application} or dynamic thresholds (DTs) \cite{wang2012collardepth,cong2022perforating}, are widely used. However, these methods rely on predefined heuristics and often fail to effectively suppress complex interference \cite{zeng2022cclsignal}. While the algorithm based on multi-hypothesis localization and Bayesian filtering demonstrate superior performance on field data \cite{seren2025probabilistic}, it necessitates multi-modal sensors. Correlation-based methods \cite{mijarez2014hpht,li2020application} are also common but are computationally intensive and suffer from limited generalization due to their reliance on fixed template matching.

Data-driven approaches have emerged as a significant research direction, demonstrating efficacy in various applications \cite{kuang2021application}, such as depth alignment \cite{torrescaceres2022automated,torrescaceres2024automated}, resistivity log interpretation \cite{noh2021deep}, perforation penetration prediction \cite{elhadidy2025optimizing}, and drilling rate evaluation \cite{liu2025research}.
Regarding collar identification, both traditional machine learning (TML) \cite{zeng2022cclsignal} and deep learning (DL) \cite{raman2024data,viggen2025improving} techniques have been utilized. While these methods offer improved generalization and automate feature extraction, they demand substantial computational resources and extensive training datasets. Currently, such processing is typically restricted to surface workstations, precluding real-time, in-situ application.
Notably, the neural network architecture proposed in \cite{raman2024data} incorporates bidirectional long short-term memory (Bi-LSTM) layers. Since Bi-LSTMs require access to future data points, they are inherently non-causal in a streaming context, rendering them unsuitable for real-time downhole deployment.
Consequently, data-driven methods require further optimization for embedded downhole systems.

Image-based recognition constitutes another category, including visual \cite{kan2020automatic,zhao2022detection,zhang2024yolo,yan2024automatic} and ultrasonic imaging \cite{viggen2025improving}. Deep neural networks have also been applied to detect casing collars within such imagery \cite{zhao2022detection,zhang2024yolo,yan2024automatic}. However, these techniques often necessitate well flushing to ensure clarity, thereby increasing operational costs. Additionally, they require reduced RIH speeds to prevent motion blur, a constraint incompatible with the requirements of wireless perforating. Furthermore, the high computational overhead and extensive training datasets requirements further render these methods impractical for autonomous downhole processing without surface support.

\begin{figure*}[!b]
    \centering
    \includegraphics[width=0.8\linewidth]{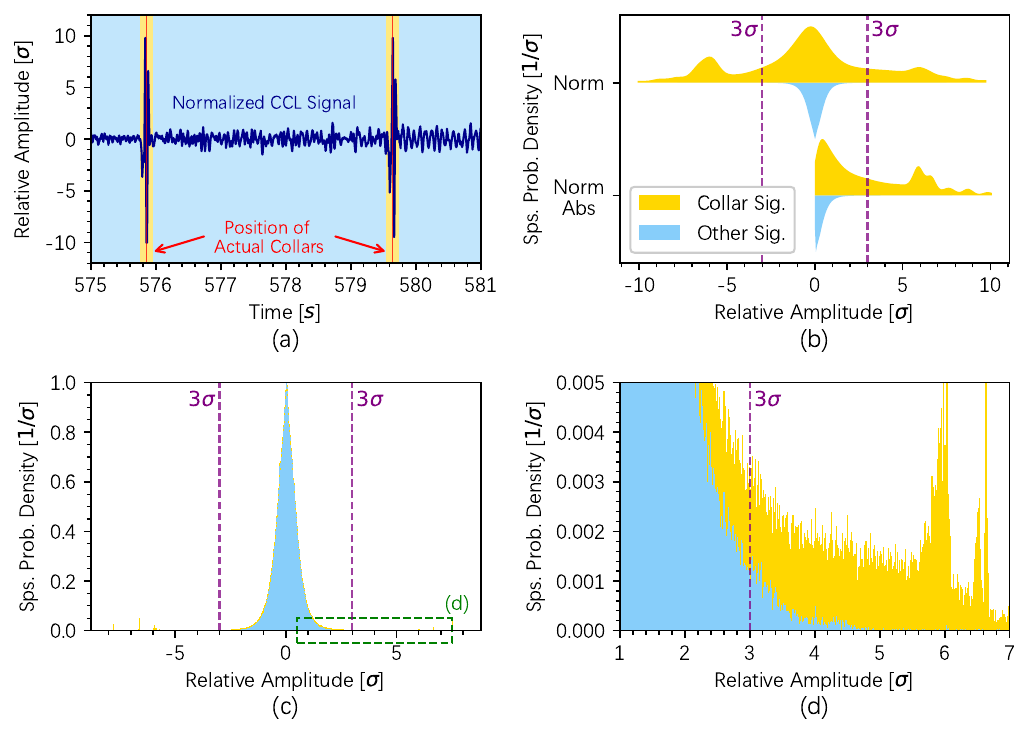}
    \caption{
        \textbf{(a)}~Normalized CCL signal with collar signals labeled in gold and non-collar signals in sky blue; all neighboring signals corresponding to actual collar locations are classified as collar signals;
        \textbf{(b)}~Violin plots of normalized samples and their absolute values for collar and non-collar signals;
        \textbf{(c)}~Probability density distributions of relative amplitudes for normalized CCL and non-collar signal samples;
        \textbf{(d)}~Enlarged view of a section from (c).
    }
    \label{fig5}
\end{figure*}

The limitations of existing methodologies highlight the need for a novel approach. An optimal solution should:
\begin{enumerate}
\item Enable precise downhole navigation and perforation at designated intervals based on collar signatures;
\item Ensure compatibility with existing downhole equipment and operational procedures, encompassing both wireline and wireless perforating approaches;
\item Perform real-time, in-situ processing of CCL signals to discriminate collar signatures from interference directly on the downhole tool;
\item Maintain a lightweight, low-cost profile suitable for deployment on embedded platforms on the downhole tool;
\item Remain robust against variations in RIH speed and eliminate the requirement for well flushing.
\end{enumerate}

~

To bridge this gap, this study proposes the Dynamic Threshold and Physical Plausibility Depth Measurement and Perforation Control (DTPPMP) system, as illustrated in Fig.~\ref{fig5}. This system is integrated into a perforating gun, a standard downhole tool. It is designed to acquire and process CCL signals, distinguish collar signatures from interference, and accurately determine depth to perforate at designated intervals.
All signals are automatically acquired and autonomously processed in situ within the perforating gun, eliminating the need for analog signal transmission over extended wirelines. This approach mitigates the signal degradation inherent to conventional methods. The in-situ identification of casing collars is enabled by a lightweight algorithm deployed on the STM32H725 microcontroller of the DTPPMP system.
By integrating the algorithm into the downhole equipment for in-situ processing, perforation operations are simplified, making the surface equipment illustrated in Fig.~\ref{fig2} unnecessary. Furthermore, operational costs associated with drilling platforms, surface infrastructure, and personnel, which constitute a significant portion of total expenses, can be substantially reduced through the use of autonomous, disposable perforating guns \cite{entchev2011autonomous}.

The DTPPMP system was validated through field trials in a vertical oil well in Sichuan, China. Experimental results demonstrate that the system accurately identifies collar signatures, performs real-time depth measurement, and executes perforation at target intervals.

~

The main contributions of this work are summarized as follows:
\begin{enumerate}
\item We propose the DTPPMP system, a cost-effective solution integrated into the perforating gun. This system enables precise, real-time, and in-situ depth measurement for perforation at designated intervals.
\item We develop a computationally efficient algorithm for CCL signal recognition and depth measurement based on correlating CCL signals with the casing tally. And we analyze the meaning and impact of parameters in the algorithm.
\item We validate the feasibility and accuracy of the DTPPMP system through field experiments and benchmark its performance against existing methods.
\end{enumerate}

This applied research paper focuses on the signal processing, depth measurement, and system design of self-locating perforating systems. The remainder of this paper is organized as follows: Section~\ref{sec:csa} presents an analysis of the CCL signal characteristics. Section~\ref{sec:proc} describes the signal processing and control algorithms, including dynamic-threshold-based collar recognition, physical plausibility verification, kinematic state calculation, and initiation determination. Section~\ref{sec:sdi} details the design and implementation of the DTPPMP system, encompassing both hardware and software architectures. Section~\ref{sec:evrs} presents the experimental methods and results. Section~\ref{sec:dis} provides a discussion of these results, including a sensitivity analysis, field test evaluations, and comparisons with existing methods. Finally, Section~\ref{sec:con} concludes the paper.

\section{CCL Signal Analysis}\label{sec:csa}

As previously established, the raw signal acquired from the CCL is an analog waveform comprising collar signatures, interference signals, and noise, as illustrated in Figs.~\ref{fig3} and~\ref{fig4}. Frequency-domain analysis reveals that both casing collar signatures and interference signals typically occupy the frequency band of \SIrange{10}{40}{\hertz}, whereas noise is predominantly concentrated in the higher frequency spectrum.
Furthermore, based on the characterization of CCL signals in \cite{song2023finite}, the temporal width of the collar signatures is inversely proportional to the logging speed, negatively correlated with the CCL coil length, and independent of the casing collar length.

Each recognizable collar signature, as shown in Fig.~\ref{fig3}, features a main peak with a significantly larger amplitude than the prevalent signals. To identify these main peaks, the CCL waveform is first normalized; subsequently, the collar signatures and non-collar signals are color-coded gold and sky-blue, respectively, as shown in Fig.~\ref{fig5}(a). To simplify the analysis, all adjacent waveform segments corresponding to actual collar locations are classified as part of the collar signatures. Because Fourier transform analysis yields suboptimal results for this differentiation, signal amplitude and statistical distribution are utilized as the primary distinguishing features between collar signatures and interference signals.

\begin{figure*}[!htbp]
    \centering
    \begin{overpic}[
        width=\linewidth,
    ]{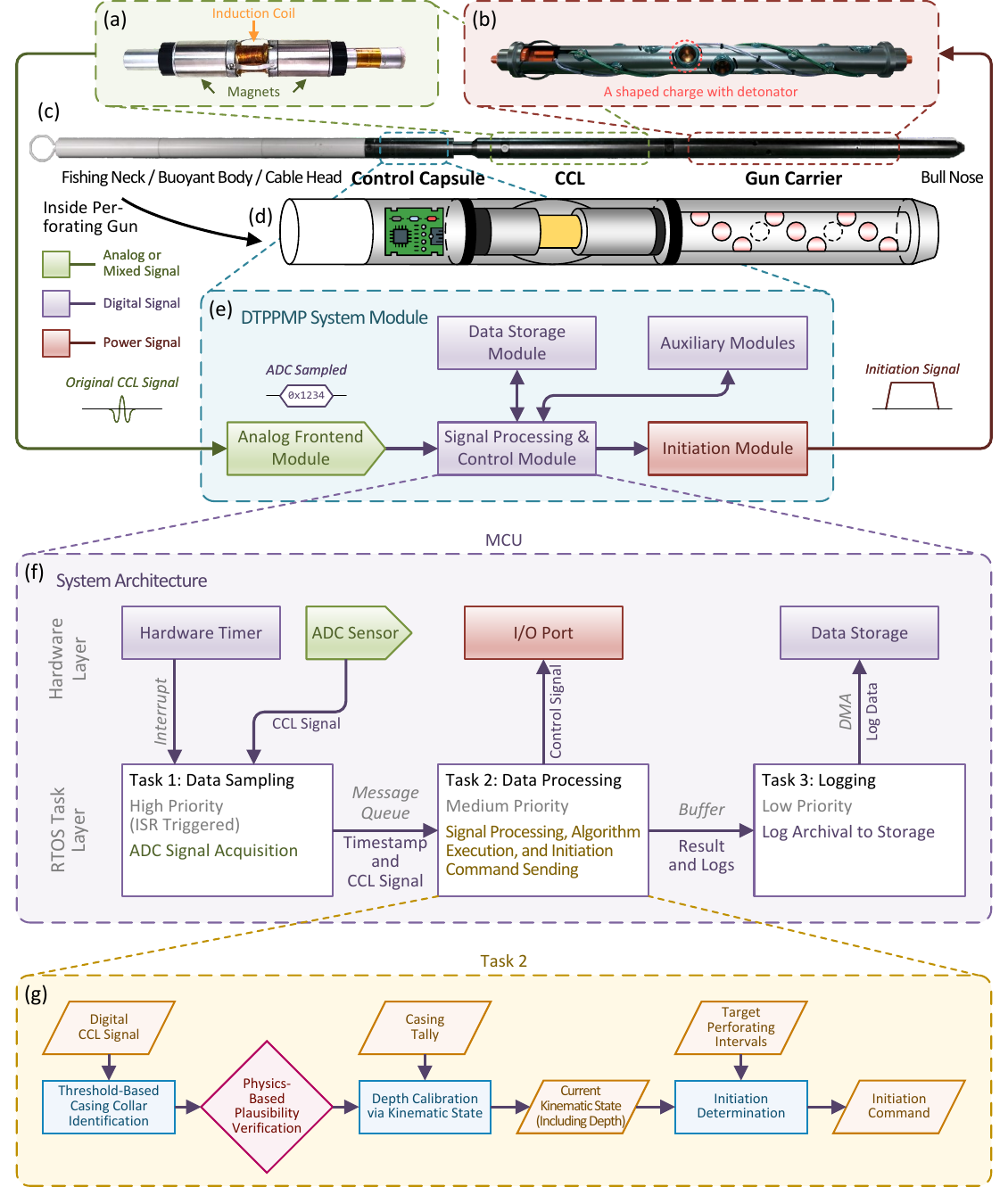}
        \fontsize{6}{6}\selectfont\sffamily
        \put(6, 2) {
            refer to Fig.~\ref{fig7}
        }
        \put(25, 2) {
            refer to Fig.~\ref{fig8}
        }
    \end{overpic}
    \caption{
        \textbf{(a)}~Structure of a casing collar locator (CCL) with the outer metal shell removed;
        \textbf{(b)}~Structure of a shaped charge carrier with the outer metal shell removed;
        \textbf{(c)}~Photograph of the actual perforating gun employed in this study;
        \textbf{(d)}~Schematic of the internal structure of the perforating gun;
        \textbf{(e)}~Functional block diagram of the DTPPMP system hardware, housed as a module within the control capsule;
        \textbf{(f)}~Block diagram of the software architecture for the firmware running on the MCU, detailing RTOS task scheduling and driving logic;
        \textbf{(g)}~Process flow diagram for casing collar identification, physical plausibility verification, kinematic state calculation, and initiation determination within the processing and control module (logging operations are omitted for clarity).
    }
    \label{fig6}
\end{figure*}

Fig.~\ref{fig5}(b) and~(c) present a violin plot and a histogram, respectively, illustrating the probability densities of the two signal types. Specifically, the violin plot in Fig.~\ref{fig5}(b) displays the absolute values of the normalized waveforms for both collar signatures and non-collar signals. To provide greater detail, a partial enlargement of Fig.~\ref{fig5}(c) is shown in Fig.~\ref{fig5}(d). As indicated by Fig.~\ref{fig5}(c), collar signatures constitute a marginal fraction of the total signal data. Furthermore, Fig.~\ref{fig5}(b) and~(d) reveals that collar signature samples predominantly fall outside three standard deviations ($3\sigma$) from the mean, a characteristic rarely exhibited by non-collar signals. These statistical characteristics suggest that collar signatures can be reliably identified from streaming data by detecting relative amplitudes that exceed a high threshold of statistical significance.

\section{Processing Algorithm}\label{sec:proc}

To ensure precise perforation based on accurate depth measurement, the overall procedure is divided into four primary stages: casing collar recognition, physical plausibility verification, kinematic state calculation, and initiation determination, as illustrated in Fig.~\ref{fig6}(g).

Initially, the casing collar recognition algorithm processes the streaming CCL signal to identify collar signature candidates, as depicted in Fig.~\ref{fig7}. These candidates are subsequently evaluated through a physical plausibility verification process; invalid candidates are discarded, as shown in Fig.~\ref{fig8}. The retained valid candidates are then utilized to calculate the kinematic state (specifically its depth, velocity, and acceleration) of the downhole toolstring by correlating them with the casing tally, which facilitates further depth calibration. Real-time depth is continuously updated based on the latest kinematic state. Finally, the initiation determination algorithm monitors this depth and issues initiation commands upon reaching the designated perforation intervals.

\subsection{Dynamic-Threshold-Based Collar Recognition}

\begin{figure}[!t]
    \centering
    \includegraphics[width=0.5\linewidth]{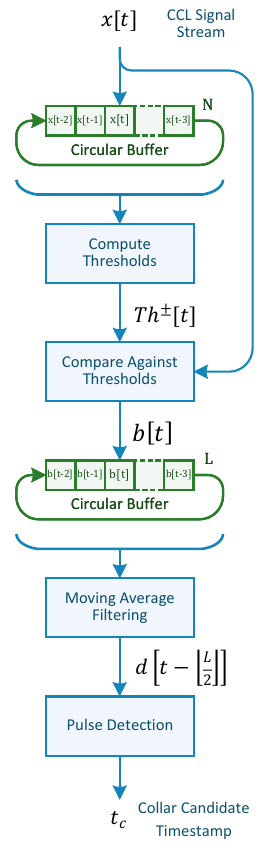}
    \caption{
        Flowchart of the casing collar recognition algorithm based on a dynamic amplitude threshold.
    }
    \label{fig7}
\end{figure}

To identify the main peaks of the collar signatures, a dynamic amplitude threshold is calculated using the standard deviation of the signal samples within a recognition window. The upper and lower thresholds are defined as:
\begin{subequations}
\begin{equation}
    {Th}^{\pm}\left[t\right] = \mu \pm \kappa\cdot\sqrt{\frac{1}{N} \sum_{i=t-N+1}^{t} \left(x\left[i\right]\right)^{2} - \mu^{2}}
    \label{eq:dyth}
\end{equation}
\begin{equation}
    \mu = \frac{1}{N} \sum_{i=t-N+1}^{t} x\left[i\right]
    \label{eq:mu}
\end{equation}
\end{subequations}
where $x\left[t\right]$ denotes the CCL signal sample at discrete time index $t$; $N$ is the width of the recognition window; and $\kappa$ is the significance coefficient.
To calculate the thresholds efficiently, the integer sample values are pushed into a circular buffer acting as a circular queue, as illustrated in Fig.~\ref{fig7}. In each iteration, the sums of $x\left[t\right]$ and $\left(x\left[t\right]\right)^{2}$ are updated using the newly acquired sample and the discarded sample, while floating-point operations are deferred to the final step. Because the raw samples are integers, this arithmetic rearrangement is exact and valid. This optimization eliminates redundant calculations, thereby reducing the time complexity from $\mathcal{O}(N)$ to $\mathcal{O}(1)$.

Signal samples falling outside the range defined by the upper and lower thresholds are considered potential collar signals, represented by a binary sequence termed the potential mask. This mask, $b\left[t\right]$, is determined as follows:
\begin{equation}
    b\left[t\right] = \begin{cases}
        0, & {Th}^{-}\left[t\right] < x\left[t\right] < {Th}^{+}\left[t\right] \\
        1, & \text{otherwise}
    \end{cases}
    \label{eq:ptmask}
\end{equation}
To eliminate glitches, the potential mask is smoothed using a moving average filter with a window width of $L$.
Omitting the initial scaling factor, the expression simplifies to:
\begin{equation}
    c\left[t\right] = \sum_{i=0}^{L-1} b\left[t - \left\lfloor\frac{L}{2}\right\rfloor + i\right]
    \label{eq:smooth}
\end{equation}
A continuous potential collar mask, $d\left[t\right]$, is then obtained via:
\begin{equation}
    d\left[t\right] = \begin{cases}
        1, & c\left[t\right] > \tau_c \\
        0, & \text{otherwise}
    \end{cases}
    \label{eq:contmask}
\end{equation}
where $\tau_c$ is a predefined threshold for the smoothed score. Through standard pulse detection, the temporal centroid of the mask $d\left[t\right]$ is extracted and identified as the collar candidate, denoted generally as $t_c$. The computational optimization for this filtering process mirrors that of the threshold calculation. It should be noted that the moving average filter inherently introduces a delay of $\left\lfloor\frac{L}{2}\right\rfloor$ samples.

\subsection{Physical Plausibility Verification and Depth Calibration}

In the worst-case scenario, the amplitude of an interference signal may be as large as that of a genuine collar signature (hereafter referred to as a ``real collar''), as illustrated in Fig.~\ref{fig4}. These misleading signals, termed ``fake collars'', must be excluded to prevent measurement inaccuracies.

Due to the large mass of the perforating gun, abrupt changes in its velocity are physically implausible. Therefore, fake collars can be excluded by calculating the average velocity associated with a collar candidate and verifying that this velocity falls within a reasonable range.

For the $k$-th valid collar, the average velocity is given by:
\begin{subequations}
    \label{eq:avgvcal}
    \begin{equation}
        \overline{V}_k = \frac{D_k-D_{k-1}}{t_k-t_{k-1}}
        \label{eq:avgvk}
    \end{equation}
    \begin{equation}
        \Delta \overline{V}_k = \overline{V}_k - \overline{V}_{k-1}
        \label{eq:davgvk}
    \end{equation}
\end{subequations}\addtocounter{equation}{-1}%
where $D_k$ and $t_k$ represent the depth and time of the $k$-th valid collar, respectively; $\overline{V}_k$ denotes the average velocity between the $(k-1)$-th and $k$-th valid collars; and $\Delta \overline{V}_k$ is the change in average velocity compared to the preceding interval. Notably, $D_k$ is derived from the casing tally, and $t_k$ is the timestamp of the verified collar.
For a collar candidate, the average velocity is calculated as:
\begin{subequations}\setcounter{equation}{2}%
    \begin{equation}
        \overline{V}_c = \frac{D_k-D_{k-1}}{t_c-t_{k-1}}
        \label{eq:avgvc}
    \end{equation}
    \begin{equation}
        \Delta \overline{V}_c = \overline{V}_c - \overline{V}_{k-1}
        \label{eq:davgvc}
    \end{equation}
\end{subequations}
where $\overline{V}_c$ represents the hypothetical average velocity if the candidate were the $k$-th valid collar, and $\Delta \overline{V}_c$ denotes the difference between this candidate's velocity and the preceding valid average velocity.

\begin{figure}[!t]
    \centering
    \begin{overpic}[
            width=1\linewidth,
        ]{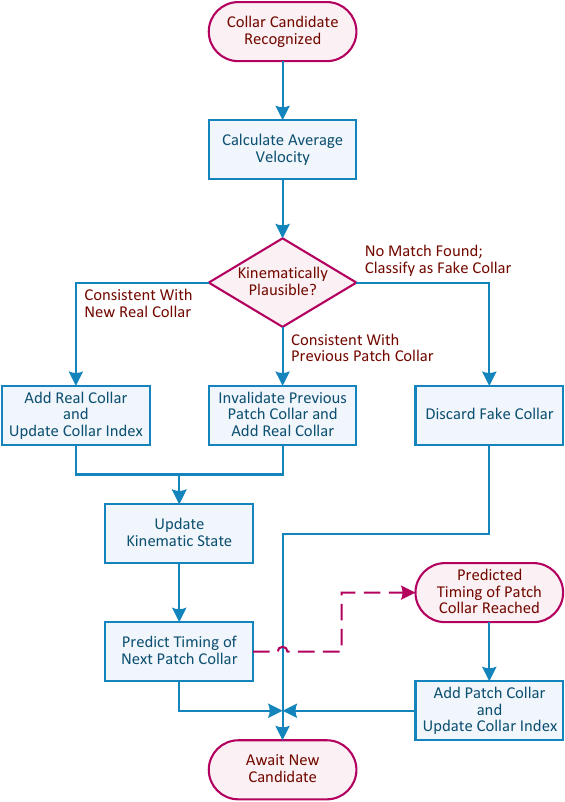}
        \fontsize{6}{6}\selectfont\sffamily
        \put(17, 81) {
            refer to (\ref{eq:avgvcal})
        }
        \put(17, 66) {
            refer to (\ref{eq:realcollar})
        }
        \put(26, 56) {
            refer to (\ref{eq:patchcollar})
        }
        \put(3, 33) {
            refer to (\ref{eq:knmdiff})
        }
        \put(3, 18) {
            refer to (\ref{eq:knmit})
        }
    \end{overpic}
    \caption{
        Flowchart of the physical plausibility verification algorithm based on physical laws and kinematic state calculations.
    }
    \label{fig8}
\end{figure}

There are three possible classifications for a collar candidate: a newly recognized real collar, a replacement for a patch collar, and a fake collar (an invalid collar), as illustrated in Fig.~\ref{fig8}. The candidate is classified as a newly recognized real collar if the following condition is satisfied:
\begin{equation}
    \left\lvert \frac{\Delta \overline{V}_c}{\Delta \overline{V}_{k-1}} \right\rvert < \tau_{dv}
    \label{eq:realcollar}
\end{equation}
where $\tau_{dv}$ denotes a fixed threshold for the relative change in the average velocity difference. When a candidate is recognized as a new real collar, its timestamp is recorded as $t_k$, and its depth is derived from the casing tally. The index for the subsequent real collar is then incremented to $(k+1)$.

If condition (\ref{eq:realcollar}) is not satisfied and the most recent collar is a patch collar, the following criterion is applied to determine whether the candidate corresponds to the patch collar predicted during the previous verification step:
\begin{equation}
    \Delta \overline{V}_c < \Delta \overline{V}_k
    \label{eq:patchcollar}
\end{equation}
where $\overline{V}_k$ is derived from the corresponding patch collar.
Upon such recognition, the candidate replaces the previous patch collar as a verified real collar, and the previous patch collar item is invalidated.

If conditions (\ref{eq:realcollar}) and (\ref{eq:patchcollar}) are not satisfied, the candidate is classified as a fake collar and subsequently discarded.

To determine the kinematic state of the downhole toolstring during the subsequent collar interval (between the latest verified collar and the next) using the timestamps and depths of latest verified collars, the differential equations governing velocity and acceleration are formulated as follows:
\begin{equation}
    \left\{
    \begin{aligned}
        \frac{\mathrm{d}}{\mathrm{d}t}D &= V \\
        \frac{\mathrm{d}}{\mathrm{d}t}V &= a \\
        \frac{\mathrm{d}}{\mathrm{d}t}a &= 0 \\
    \end{aligned}
    \label{eq:knmdiff}
    \right.
\end{equation}
where $D$, $V$, $a$, and $t$ represent the depth, velocity, acceleration, and time, respectively. To simplify the model, it is assumed that the forces acting between consecutive collars remain unchanged, resulting in a constant acceleration.
The corresponding boundary conditions are given by:
\begin{equation}
    \left\{
    \begin{aligned}
        D\left[t_1\right] &= D_1 \\
        D\left[t_2\right] &= D_2 \\
        D\left[t_3\right] &= D_3 \\
    \end{aligned}
    \right.
    \label{eq:knmbc}
\end{equation}
where $t_1$, $t_2$, and $t_3$ represent the timestamps of three consecutive real collars (note that this indexing differs from the definition of $t_k$); and $D_1$, $D_2$, and $D_3$ represent their corresponding depths. Assuming $t_3$ is the most recent timestamp, the remaining kinematic state of this collar, denoted by $V_3$ and $a_3$, can be determined. Subsequently, the depth and velocity at any given time $t$ are calculated using the following equations:
\begin{equation}
    \left\{
    \begin{aligned}
        D\left[t\right] & = D_3 + V_3 \left(t - t_3\right) + \frac{1}{2} a_3 \left(t - t_3\right)^2 \\
        V\left[t\right] & = V_3 + a_3 \left(t - t_3\right)                                          \\
    \end{aligned}
    \right.
    \label{eq:knmit}
\end{equation}
The equation coefficients are updated upon the verification of a new real collar.

By substituting the depth of the next real collar, $D_{k+1}$, the arrival time of the next real collar can be predicted using the following equation:
\begin{equation}
    D_{k+1} = D_k + V_k \left(t_p - t_k\right) + \frac{1}{2} a_k \left(t_p - t_k\right)^2
    \label{eq:tppred}
\end{equation}
where $t_p$ represents the predicted timestamp of the next real collar. The positive root of the quadratic equation is selected to ensure that $t_p > t_k$.

If a collar candidate does not appear at the predicted time, which may be due to missed detection or unexpected deceleration, a patch collar is inserted at the predicted timestamp, as shown in Fig.~\ref{fig8}. In a deceleration scenario, the patch collar is invalidated once the corresponding real collar is detected; in a missed detection scenario, adding a patch collar prevents indexing misalignments within the casing tally.

In summary, physical plausibility verification is applied to correct potential errors in casing collar identification, thereby improving the overall accuracy of the depth measurement.

\subsection{Initiation Determination}

Based on (\ref{eq:knmit}), the initiation time is calculated using the latest kinematic state and the depth of the target interval. To ensure safety, the initiation time is established only if the most recently detected collar is verified as a real collar. Additionally, a software lockout mechanism based on depth and time is implemented to prevent accidental initiation.

\section{System Design and Implementation}\label{sec:sdi}

\subsection{System Architecture}

The DTPPMP system is integrated into a perforating gun, as illustrated in Fig.~\ref{fig6}. The perforating gun comprises three primary components: a control capsule, a CCL, and a gun carrier serving as the payload, as shown in Fig.~\ref{fig6}(c) and (d).
Unlike typical perforating gun structures, the gun carrier, on which the shaped charges are installed, is positioned at the head rather than the middle, as shown in Fig.~\ref{fig6}(b).
The CCL is located in the middle and is identical to the conventional Type CNLC-73-150 model. Its structure is depicted in Fig.~\ref{fig6}(a). The control capsule, situated at the tail, contains the system module, battery, and top sub-assembly. For wireline perforating, the top sub-assembly is equipped with a cable head; for wireless perforating, it features a buoyant body or a fishing neck for recovery operations. The key hardware specifications of the DTPPMP system are summarized in Table~\ref{tab:1}.

\begin{table*}[!b]
    \caption{
        Key Hardware Specifications of the DTPPMP System
    }
    \label{tab:1}
    \centering
    \renewcommand{\arraystretch}{1.2}
    \resizebox{\textwidth}{!}{
        \begin{tabularx}{\textwidth}{>{\centering\arraybackslash}l l X}
\toprule
    \textbf{Component}                 &
    \textbf{Model / Type}              &
    \textbf{Specifications and Description}                                                                                            \\
\midrule
    MCU                                &
    STM32H725                          &
    ARM Cortex-M7 core, \SI{550}{\mega\hertz}, handling signal processing and logic control.                                           \\
    ADC                                &
    Built-in ADC of STM32H7            &
    \SI{16}{\bit} resolution, Max. \SI{3.6}{\mega\sample\per\second} sampling rate.                                                         \\
    AFE                                &
    Proposed Design (Fig.~\ref{fig9})  &
    Low-noise, rail-to-rail operational amplifier for signal conditioning and gain adjustment.                                         \\
    Storage                            &
    W25Q02                             &
    \SI{2}{\giga\bit} capacity for logging raw CCL data and system logs.                                                               \\
    Power Supply                       &
    High-temp Li-SOCl\textsubscript{2} &
    \SI{12}{\volt}, \SI{14}{\ampere\hour}, Max.\ Temp.\ \SI{150}{\degreeCelsius}, designed for downhole high-temperature environments. \\
    CCL Sensor                         &
    Type CNLC-73-150                   &
    $R_{\mathrm{DC}}$ = \SI{180}{\ohm}, \num{6000}~turns, SmCo magnet (\SI{2000}{\gauss})                                              \\
\bottomrule
        \end{tabularx}
    }
\end{table*}

\subsection{System Module Overview}

As depicted in Fig.~\ref{fig6}(e), the system module housed within the control capsule comprises several submodules, including an analog front-end (AFE) module, a signal processing and control module, a data storage module, an initiation module, and other auxiliary modules (e.g., an I/O module).
The signal processing and control module is built around the STM32H725, a high-performance microcontroller unit (MCU) featuring a 32-bit ARM Cortex-M7 core operating at a clock speed of \SI{550}{\mega\hertz}. The MCU governs the overall operation of the system, including the acquisition and processing of raw digitized CCL signals, collar identification, depth measurement, and initiation module control. Additionally, the STM32H725 features two integrated 16-bit analog-to-digital converters (ADCs) with a maximum sampling rate of \SI{3.6}{\mega\sample\per\second}.

\subsection{AFE Module}

\begin{figure}[!b]
    \centering
    \includegraphics[width=0.9\linewidth]{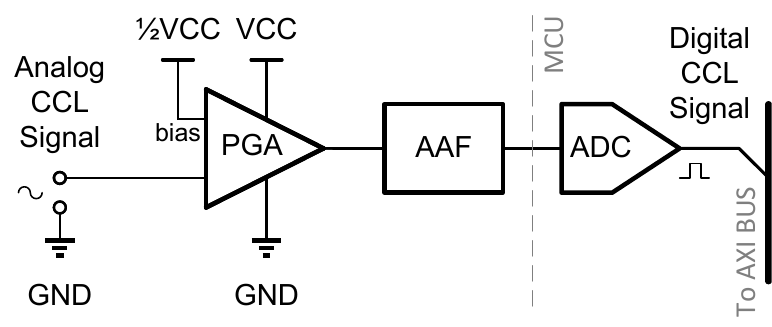}
    \caption{
        Schematic of the analog front-end (AFE) module within the DTPPMP system, comprising a programmable gain amplifier (PGA), an anti-aliasing filter (AAF), and an analog-to-digital converter (ADC).
    }
    \label{fig9}
\end{figure}

As the perforating gun descends into the well via a winch or gravity, the AFE module acquires the analog CCL signals and converts them into a digital format. As detailed in Fig.~\ref{fig9}, the AFE comprises a programmable gain amplifier (PGA), an anti-aliasing filter (AAF), and the integrated ADC of the MCU. The PGA scales the signal amplitude to match the required dynamic range. The AAF acts as a low-pass filter to attenuate high-frequency noise.
The ADC digitizes the conditioned CCL signal at a sampling rate of \SI{1}{\kilo\hertz} with 16-bit resolution, and the resulting digital stream is transferred to the MCU core for subsequent processing. Given the low-frequency characteristics of the signals analyzed previously, this streamlined hardware architecture is sufficient.

\subsection{Signal Processing and Control Module}

The collar identification and depth measurement algorithms, executed by the MCU, identify genuine collar signatures, filter out interference, and calculate the kinematic state (including descent depth) in real time, as illustrated in Fig.~\ref{fig6}(g). Upon reaching the target perforation intervals, the initiation module triggers the detonation of the shaped charges, as shown in Fig.~\ref{fig1}(b) and~(c). Log data generated during sampling and algorithmic processing are stored in the data storage module. Following the perforation operations, the control capsule is recovered.

To realize these functionalities, the firmware is designed to fully utilize the platform's capabilities, as illustrated in Fig.~\ref{fig6}(f). FreeRTOS, a real-time operating system (RTOS), manages system tasks, including sampling, processing, and logging. Triggered by a hardware timer interrupt to ensure precise sampling intervals, the sampling task acquires the digitized CCL data from the ADC and transmits the corresponding timestamps and ADC values to the processing task via a message queue. Upon receiving this message, the processing task executes the identification algorithms and caches the resulting log data in a buffer. Once the buffer reaches capacity, the logging task writes the data to the storage module via direct memory access (DMA).

\subsection{Workflow of the DTPPMP System}

Based on the system design, the operational workflow of the DTPPMP system proceeds as follows:
\begin{enumerate}
\item System Initialization: This phase includes the assembly of the perforating gun, configuration input, and pre-deployment surface checks.
\item Activation: The DTPPMP system is activated upon entry into the wellbore. For wireline perforating, the gun is lowered via a winch; for wireless perforating, it may descent by gravity.
\item Depth Measurement: The system recognizes casing collar signatures using physical plausibility verification and calculates the kinematic state to continuously calibrate depth measurements as the gun descends. Notably, unlike traditional methods that perform collar identification and perforation during the ascent, the proposed method executes these processes during the descent.
\item Initiation: The system initiates the detonation of the shaped charges upon reaching the target perforation interval.
\item Recovery: The perforating gun is retrieved from the wellbore. In wireline systems, retrieval is performed using a winch, whereas in wireless systems, the gun is recovered utilizing a fishing neck or buoyancy mechanisms.
\end{enumerate}
This complete workflow is illustrated in Fig.~\ref{fig2}.

\section{Evaluation and Results}\label{sec:evrs}

\subsection{Parameter Selection}

To determine the optimal parameters for the casing collar recognition and physical plausibility verification algorithms, historical logging data are utilized to conduct a sensitivity analysis and an ablation study. To rigorously evaluate the effectiveness of the proposed methods, logs exhibiting interference are specifically selected. Parameter sensitivity is assessed by varying a single parameter while keeping the others constant.

The recognition results are compared against the casing tally, which serves as the ground truth. Specifically, recognized collar signatures that align with the casing tally are defined as true positives (TPs); recognized collars that do not match the casing tally are defined as false positives (FPs); and undetected actual collars are defined as false negatives (FNs). Accuracy, precision, recall, and the F1-score are subsequently calculated to evaluate the performance of the DTPPMP system.

\begin{figure}[!b]
    \centering
    \includegraphics[width=1\linewidth]{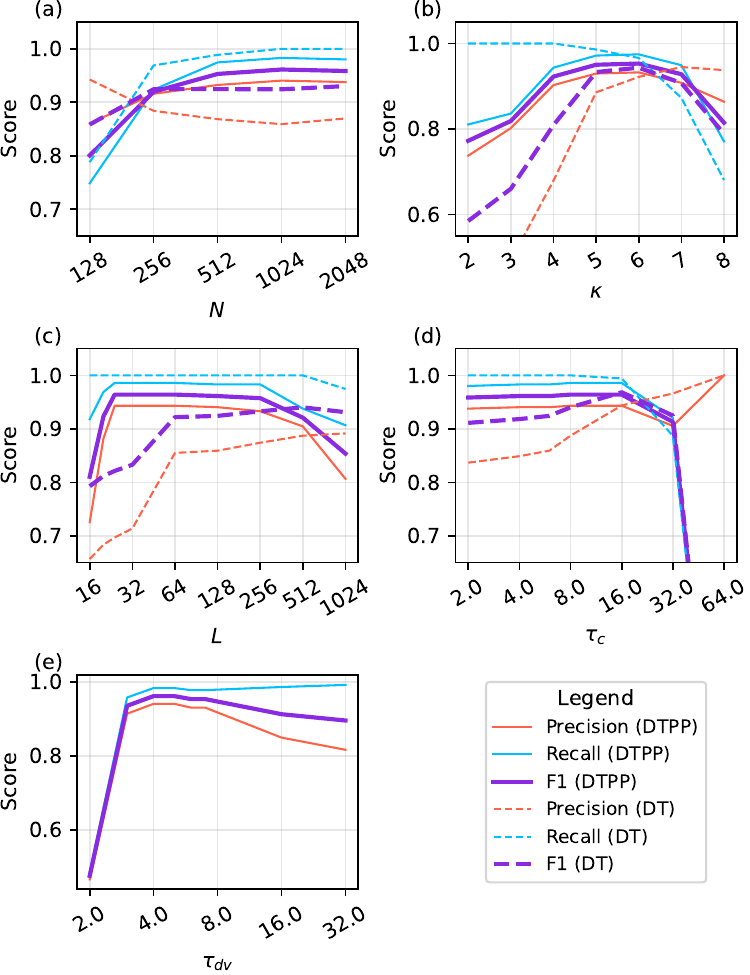}
    \caption{
        Results of the sensitivity analysis and ablation experiments:
        \textbf{(a)}~$N$: width of the recognition window;
        \textbf{(b)}~$\kappa$: significance coefficient;
        \textbf{(c)}~$L$: moving average filter window width;
        \textbf{(d)}~$\tau_c$: smoothed potential collar score threshold;
        \textbf{(e)}~$\tau_{dv}$: threshold for the relative change in the average velocity difference.
        \\
        ``DT'' and ``PP'' denote ``dynamic threshold'' and ``physical plausibility'', respectively.
    }
    \label{fig10}
\end{figure}

The evaluation results are presented in Fig.~\ref{fig10}, which compares the performance of the standalone dynamic threshold (DT) collar recognition against the DT method combined with physical plausibility verification (DTPP). Based on these experimental results, the system parameters are configured as follows:
\begin{equation}
    \begin{aligned}
        N         & = 512 \\
        \kappa    & = 5   \\
        L         & = 32  \\
        \tau_{c}  & = 8   \\
        \tau_{dv} & = 4   \\
    \end{aligned}\label{eq:params}
\end{equation}

\subsection{Field Test}

\begin{figure}[!htbp]
    \centering
    \subfloat[]{\includegraphics[width=0.6\columnwidth]{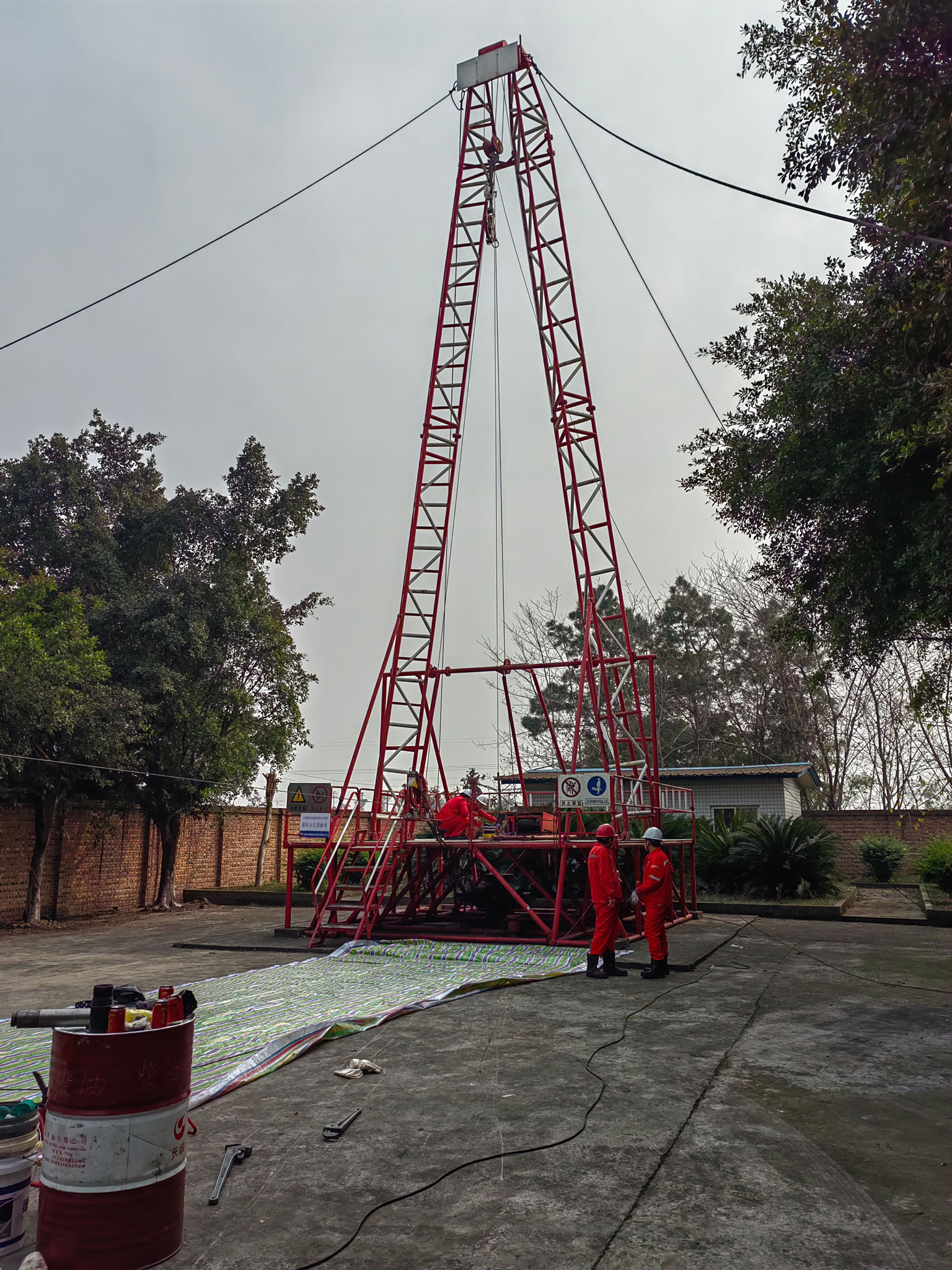}\label{fig11a}}
    \linebreak
    \subfloat[]{\includegraphics[width=0.8\columnwidth]{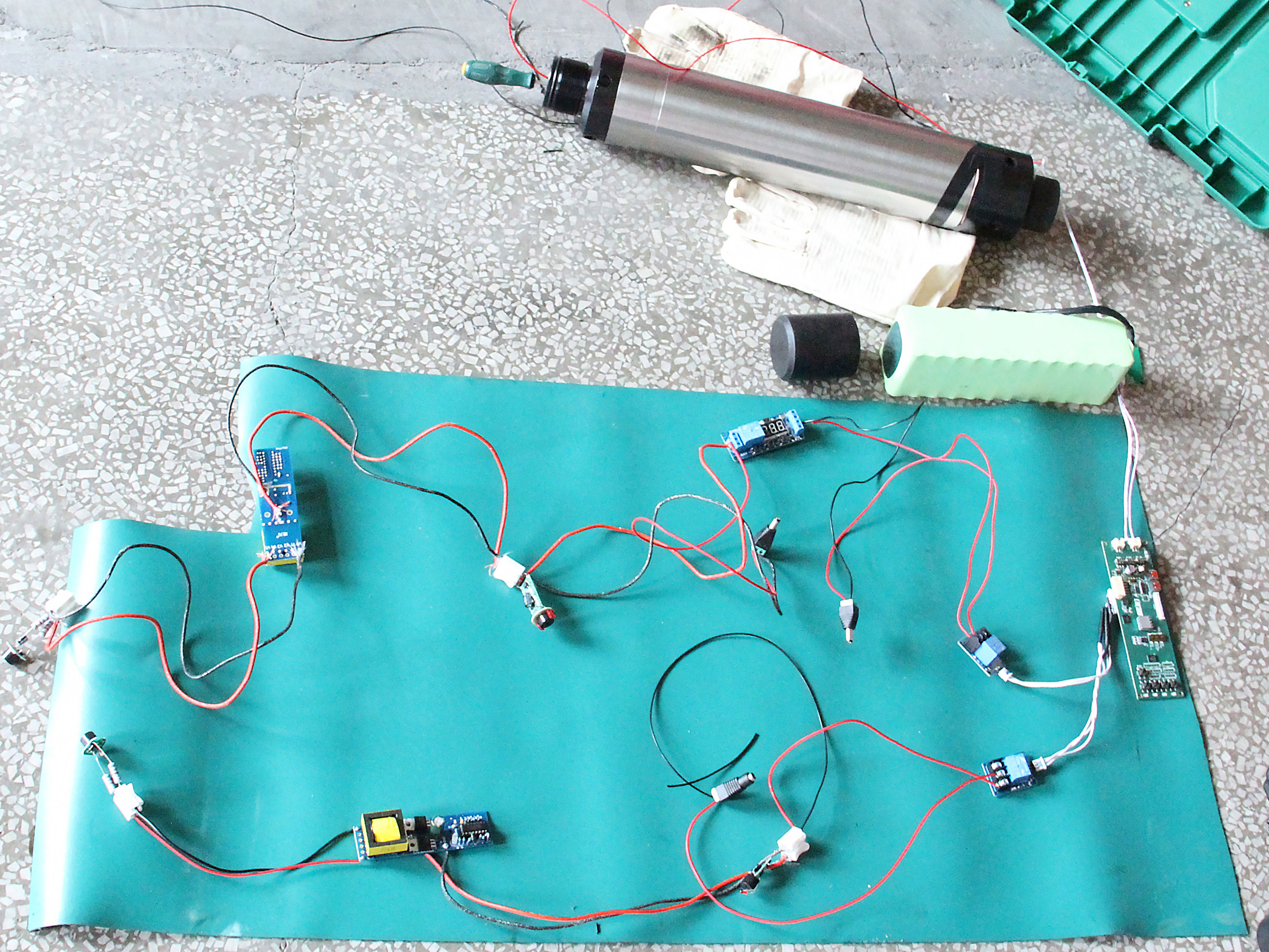}\label{fig11b}}
    \caption{
        \textbf{(a)}~Field testing of the DTPPMP system in an operational oil well;
        \textbf{(b)}~Pre-deployment check of the DTPPMP system prior to the downhole experiment (the power supply is not yet connected).
    }
    \label{fig11}
\end{figure}

An operational oil well in Sichuan, China, depicted in Fig.~\ref{fig11}(a), was utilized to verify the feasibility and reliability of the DTPPMP system. Prior to the downhole experiment, a comprehensive pre-deployment check was conducted on the surface, as illustrated in Fig.~\ref{fig11}(b).

During the downhole experiment, the perforating gun equipped with the DTPPMP system was deployed into the wellbore via a winch from the drilling platform, and the system was activated upon entry. The descent velocity accelerated from \num{0} to a constant speed of \SI{8}{\kilo\meter\per\hour}. The target perforating interval was set at \SI{1100}{\meter} from the wellhead. To ensure operational safety, live shaped charges were omitted; only detonators were installed. Following the test, the tool string was retrieved to the surface for data extraction. The entire operational sequence was recorded by the onboard storage module and subsequently exported to a computer for analysis. Such experimental procedure was repeated multiple times to ensure consistency. The data log from a representative experiment is presented in Fig.~\ref{fig12}.

In this evaluation, the recognition results were compared against the casing tally to quantify the occurrences of false positives and false negatives. Accuracy, precision, recall, and the F1-score were subsequently calculated to comprehensively assess the performance of the DTPPMP system. The resulting confusion matrix and the corresponding evaluation metrics are detailed in Table~\ref{tab:2}.

\section{Discussion}\label{sec:dis}

\subsection{Sensitivity Analysis and Ablation Study}

Fig.~\ref{fig10}(a) illustrates the sensitivity of the detection metrics, including precision, recall, and the F1-score, to the width of the recognition window, $N$. The results indicate that the system maintains stable performance and a high F1-score when $N$ exceeds \num{256}, demonstrating the robustness of the proposed method against variations in $N$. However, $N$ cannot be arbitrarily large. First, because the data within the recognition window must be buffered, a larger $N$ inherently demands increased memory consumption. Second, an excessively large $N$ causes large-amplitude signals to persist for a longer duration, potentially obscuring the recognition of small-amplitude collar signatures. Therefore, $N$ is set to \num{512}.

Fig.~\ref{fig10}(b) illustrates the sensitivity of the detection metrics to the significance coefficient, $\kappa$. The results indicate that the system maintains stable performance and a high F1-score for $\kappa$ values ranging from \numrange{4}{7}. Considering the definition of $\kappa$ alongside the statistical distributions presented in Fig.~\ref{fig5}(d), $\kappa$ is fundamentally constrained by the statistical characteristics of the signal. Specifically, an overly large $\kappa$ leads to missed detections, whereas a small $\kappa$ increases the likelihood of incorrectly identifications. Therefore, $\kappa$ is set to \num{5}.

\begin{table}[!t]
    \centering
    \caption{
        Casing Collar Recognition Performance
    }
    \label{tab:2}
    \setlength{\tabcolsep}{2pt}
    \renewcommand{\arraystretch}{1.2}
    \begin{threeparttable}
    \begin{tabular}{l @{\hspace{8pt}} rrrrr @{\hspace{8pt}} rrrr}
    \toprule
        \makecell[c]{\textbf{~}} &
        \makecell[c]{\textbf{TP}\tnote{a}} & \makecell[c]{\textbf{FP}} & \makecell[c]{\textbf{FN}} & \makecell[c]{\textbf{TN}} & \makecell[c]{\textbf{Total}\tnote{b}} &
        \makecell[c]{\textbf{Acc}} & \makecell[c]{\textbf{P}} & \makecell[c]{\textbf{R}} & \makecell[c]{\textbf{F1}} \\
    \midrule
        DTPPMP & 579 & 7 & 9 & 0 & 588 & 97.3\% & 98.8\% & 98.5\% & 98.6\% \\
    \bottomrule
    \end{tabular}
    \begin{tablenotes}\footnotesize
        \item[a] The abbreviations ``TP'', ``FP'', ``FN'', ``TN'', ``Acc'', ``P'', ``R'', ``F1'' denote true positive, false positive, false negative, true negative, accuracy, precision, recall, and F1-score, respectively.
        \item[b] The total number of ground truth casing collars.
    \end{tablenotes}
    \end{threeparttable}
\end{table}

Fig.~\ref{fig10}(c) illustrates the sensitivity of the detection metrics to the moving average filter window width, $L$. The system maintains stable performance and a high F1-score when $L$ ranges from \numrange{24}{256}, confirming the method's robustness to variations in $L$. Similar to the constraints on $N$, $L$ cannot be excessively large. Furthermore, a large $L$ introduces significant latency into the system, which is detrimental to real-time applications. Therefore, $L$ is set to \num{32}.

Fig.~\ref{fig10}(d) illustrates the sensitivity of the detection metrics to the smoothed potential collar score threshold, $\tau_c$. The performance remains stable with a high F1-score for $\tau_c$ values between \numrange{2}{16}, indicating robustness against variations in this parameter. The RIH speed influences the temporal distribution of the smoothed potential collar scores generated by the collar signatures. Consequently, a high $\tau_c$ threshold results in missed detections at elevated RIH speeds, whereas a low $\tau_c$ threshold incorrectly classifies interference as valid collar signatures. As a result, $\tau_c$ is set to \num{8}.

Finally, Fig.~\ref{fig10}(e) illustrates the sensitivity of the detection metrics to the threshold for the relative change in the average velocity difference, $\tau_{dv}$. The system exhibits stable performance and a high F1-score when $\tau_{dv}$ ranges from \numrange{3}{6}, verifying its robustness to variations in $\tau_{dv}$. Because $\tau_{dv}$ represents the permissible degree of velocity fluctuation, its optimal value is physically constrained by the mass of the perforating gun and the prevailing wellbore conditions. An excessively high $\tau_{dv}$ diminishes the system's ability to filter out false positives, while an excessively low value mistakenly rejects correct identifications. Therefore, $\tau_{dv}$ is set to \num{4}.

~

\begin{figure*}[!t]
    \centering
    \includegraphics[width=\linewidth]{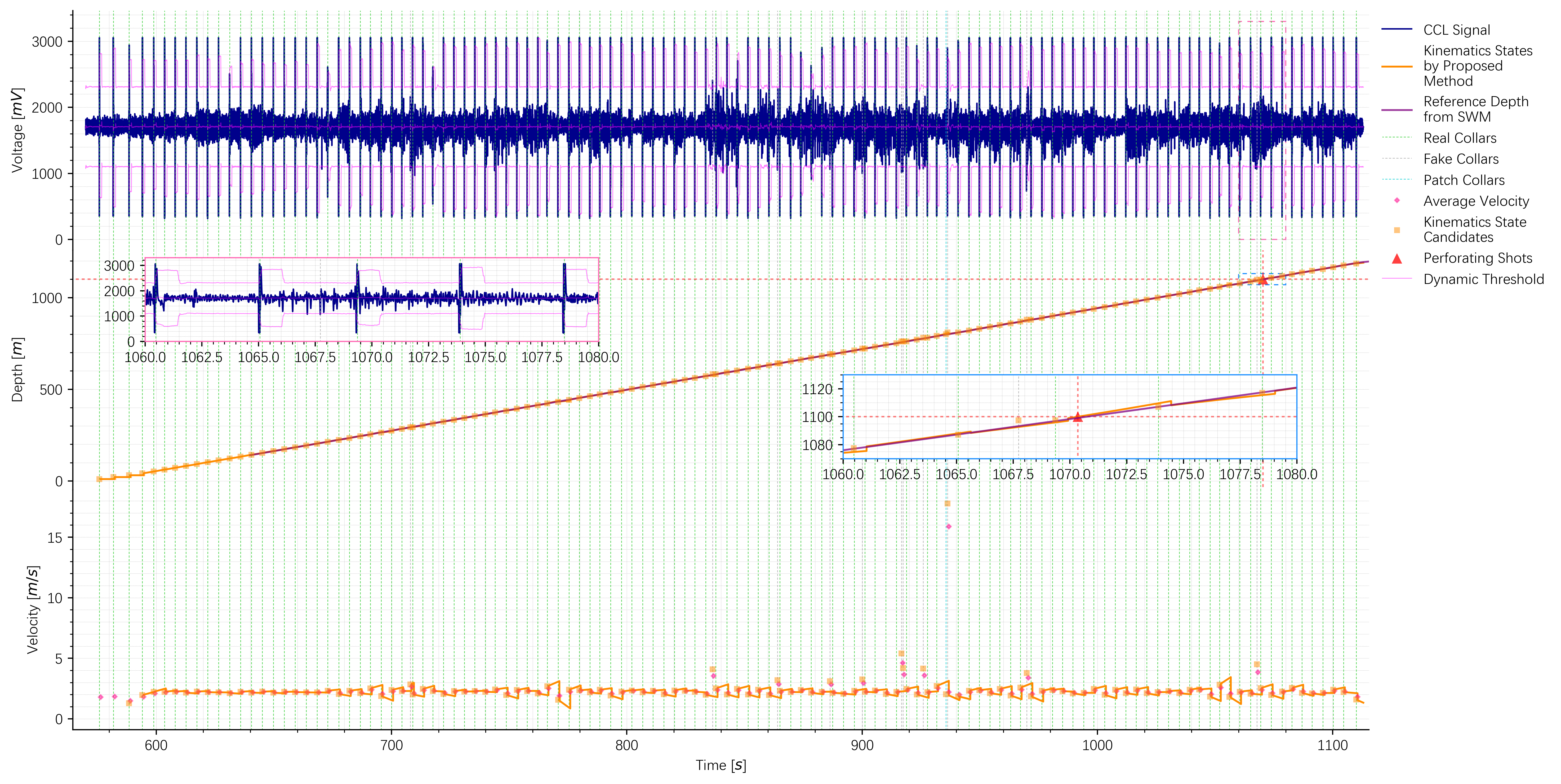}
    \caption{
        Data record from a single field test, comparing the depth derived from the CCL signal with that obtained directly from the SWM signal.
        \\
        The dynamic threshold collar recognition algorithm successfully identifies the casing collar signatures, while the physical plausibility verification algorithm eliminates incorrectly identifications and corrects missed detections.
    }
    \label{fig12}
\end{figure*}

To validate the contribution of each module, the results of the ablation study are illustrated in Fig.~\ref{fig10}(a)--(d). It can be observed that, within a reasonable range of parameter values, the physical plausibility verification algorithm improves the recognition F1-score by approximately \SIrange{3}{5}{\percent}. This demonstrates that the physical plausibility verification algorithm is effective in filtering out false positives and mitigating false negatives.

Interestingly, the overall recognition performance coupled with physical plausibility verification may even degrade compared to the baseline without physical plausibility verification if the initial performance of the dynamic threshold recognition algorithm is suboptimal. This indicates that the physical plausibility verification algorithm relies on the preceding algorithm achieving a sufficiently high baseline performance to yield further improvements. Therefore, optimizing the primary collar recognition algorithm remains a necessary prerequisite.

\subsection{Complexity Analysis}

We analyze the computational complexity of the proposed algorithm across four primary stages: casing collar recognition, physical plausibility verification, kinematic state calculation, and initiation determination. Casing collar recognition involves summation operations that are optimized using a circular buffer, resulting in $\mathcal{O}(1)$ time and $\mathcal{O}(N)$ space complexities. Physical plausibility verification computes the average velocity, which is cached in memory, similarly requiring $\mathcal{O}(1)$ time and $\mathcal{O}(N)$ space. The kinematic state calculation relies on solving a quadratic equation, taking $\mathcal{O}(1)$ time and $\mathcal{O}(1)$ space. Finally, initiation determination is a constant-time operation requiring $\mathcal{O}(1)$ time and $\mathcal{O}(1)$ space. Consequently, the overall time complexity of the proposed algorithm is $\mathcal{O}(1)$ per sample, and the space complexity is $\mathcal{O}(N)$.

When deployed on an embedded platform during our experiments, the average processing time for a single sample is \SI{1.5}{\micro\second}, which comfortably satisfies the requirements for real-time applications operating with a \SI{1}{\milli\second} sampling period. This substantial processing margin demonstrates that the proposed algorithm is highly efficient and suitable for deployment on platforms with limited computing resources. Consequently, it facilitates a reduction in overall system costs, a critical consideration for disposable applications. This computational advantage is sufficient to offset the slight shortcomings observed in the recognition metrics.

\subsection{Field Test Analysis}

Furthermore, the experimental log data were analyzed. As depicted in the experimental record in Fig.~\ref{fig12}, manual analysis of the CCL logs confirmed that the initiation signal was transmitted between the 110th and 111th casing collars, located at \SI{1097.47}{\meter} and \SI{1107.06}{\meter} from the casing tally, respectively. This corresponds to the designated perforating interval, demonstrating an error margin of less than one collar length. These field test results verify that the system operated accurately and efficiently during the experiment.

The detailed procedures of the remaining experiments are omitted for brevity; however, their comprehensive performance statistics are summarized in Table~\ref{tab:2}.

~

\begin{table*}[!b]
    \centering
    \caption{
        Performance Comparison of the DTPPMP System with Existing Methods
    }
    \label{tab:3}
    \renewcommand{\arraystretch}{1.2}
    \resizebox{\textwidth}{!}{
        \begin{threeparttable}
            \newcommand{\cdash}{%
                \multicolumn{1}{c}{--}%
            }
            \begin{tabular}{c rrrrr lcc}
\toprule
    \multicolumn{1}{c}{\multirow{2}{*}{~}}                                                      &
    \multicolumn{5}{c}{\multirow{1}{*}{\textbf{Classification Metrics}}}                        &
    \multicolumn{1}{c}{\multirow{2}{*}{\textbf{Methods and Description}}}                       &
    \multicolumn{1}{c}{\multirow{2}{*}{\makecell[c]{\textbf{In-situ \&}\\\textbf{Real-time}}}}  &
    \multicolumn{1}{c}{\multirow{2}{*}{\textbf{Method Status}}}                                 \\
\cline{2-6} &
    \multicolumn{1}{c}{\textbf{Test}}      &
    \multicolumn{1}{c}{\textbf{Acc}}    &
    \multicolumn{1}{c}{\textbf{P}}      &
    \multicolumn{1}{c}{\textbf{R}}      &
    \multicolumn{1}{c}{\textbf{F1}}     &
    ~ & ~ & ~ \\
\midrule
    {DTPPMP}                         & 588    & 97.3\%  & 98.8\%  & 98.5\% & 98.6\% & CCL + Dynamic threshold + Physical plausibility & \checkmark & Field Test  \\
\midrule
    {\cite{zeng2022cclsignal}}       & \cdash & 99.7\%  & 99.7\%  & 99.7\% & 99.7\% & CCL + (TML) Random Forest (RF)                  & ~          & Simulation  \\
    {\cite{zeng2022cclsignal}}       & \cdash & 99.6\%  & 99.6\%  & 99.6\% & 99.6\% & CCL + (TML) Adaboost                            & ~          & Simulation  \\
    {\cite{zeng2022cclsignal}}       & \cdash & 99.6\%  & 99.6\%  & 99.6\% & 99.6\% & CCL + (TML) eXtreme Gradient Boosting (XGBoost) & ~          & Simulation  \\
    {\cite{zeng2022cclsignal}}       & \cdash & 99.5\%  & 99.5\%  & 99.5\% & 99.5\% & CCL + (TML) Support Vector Machine (SVM)        & ~          & Simulation  \\
    {\cite{raman2024data}}           & 109    & \cdash  & 98.9\%  & 98.0\% & 98.5\% & CCL + (DL) BiLSTM                               & ~          & Field Test  \\
    {\cite{jing2025identification}}  & 269    & 97.8\%  & 95.9\%  & 99.1\% & 97.5\% & CCL + (DL) CNN-LSTM                             & ~          & Simulation  \\
    {\cite{jing2025identification}}  & 269    & 97.4\%  & 100.0\% & 94.2\% & 97.0\% & CCL + (DL) CNN                                  & ~          & Simulation  \\
    {\cite{raman2024data}}           & 109    & \cdash  & 93.6\%  & 97.9\% & 95.7\% & CCL + Wavelet Transform + (DL) FNN              & ~          & Field Test  \\
    {\cite{raman2024data}}           & 109    & \cdash  & 94.0\%  & 96.3\% & 95.1\% & CCL + (DL) CNN                                  & ~          & Field Test  \\
    {\cite{jing2025identification}}  & 269    & 94.8\%  & 100.0\% & 88.4\% & 93.9\% & CCL + (DL) LSTM                                 & ~          & Simulation  \\
    {\cite{wang2012collardepth}}     & 8      & 100.0\% & \cdash  & \cdash & \cdash & CCL + (DL) Relative Amplitude                   & ~          & Field Test  \\
    {\cite{mijarez2014hpht}}         & \cdash & \cdash  & \cdash  & \cdash & \cdash & CCL + Cross Correlation + Predefined Threshold  & ~          & Field Test  \\
    {\cite{li2013casing}}            & \cdash & \cdash  & \cdash  & \cdash & \cdash & CCL + Wavelet Transform                         & ~          & Theoretical \\
\midrule
    {\cite{zhang2024yolo}}           & 110    & 99.5\%  & 99.8\%  & 99.7\% & 99.7\% & VideoLog + (DL) YOLOv5                          & ~          & Simulation  \\
    {\cite{yan2024automatic}}        & 67     & 99.0\%  & \cdash  & \cdash & \cdash & VideoLog + (DL) Faster-RCNN                     & ~          & Simulation  \\
    {\cite{kan2020automatic}}        & 103    & 86.9\%  & \cdash  & \cdash & \cdash & VideoLog + Graphical Features                   & ~          & Simulation  \\
\bottomrule
            \end{tabular}
            \begin{tablenotes}\footnotesize
\item ``--'' indicates ``Not reported'' or ``Not applicable''; ``Acc'', ``P'', ``R'', ``F1'' denote accuracy, precision, recall, and F1-score, respectively.
\item Note that the performance metrics of existing works are cited directly from their respective original papers; consequently, methodological paradigms and experimental conditions (e.g., datasets and evaluation methodologies) may vary.
            \end{tablenotes}
        \end{threeparttable}
    }
\end{table*}

Table~\ref{tab:3} presents a comparison between the recognition performance of the proposed DTPPMP and that of existing methods. Within the category of non-machine learning approaches, DTPPMP consistently achieves top-tier performance in terms of both accuracy and F1-score. Although methods utilizing neural networks yield an improvement of approximately \SI{2}{\percent} in F1-score, they demand significantly higher computational resources. Furthermore, the reported F1-score of DTPPMP surpasses the metrics reported for the neural network-based CCL signal identification methods reported in \cite{raman2024data,jing2025identification}. Significantly, among the methods listed in Table~\ref{tab:3}, DTPPMP is the only approach capable of performing real-time, in-situ recognition and depth measurement.

However, establishing a direct, unified benchmark presents critical challenges due to the diversity of methodological paradigms in the literature, the frequent unavailability of algorithmic implementation details, and the inaccessibility of proprietary datasets utilized in related works. Table~\ref{tab:3} systematically compiles the available reported performance metrics of existing methods, thereby providing a comparative overview. Even when accounting for variations in experimental setups, the proposed DTPPMP system exhibits distinct advantages in recognition performance, parameter efficiency, and computational economy relative to the existing literature.

\subsection{Effects of Operation Configurations}

Although the field validation was conducted on a specific well, the proposed method is designed based on the fundamental physical characteristics of casing collars. As discussed in \cite{song2023finite}, the effects of various well conditions and operational parameters are summarized as follows:
\begin{itemize}
\item \textit{Casing inner diameter:} Inversely affects the amplitude of the collar signature.
\item \textit{Casing thickness:} Exhibits no significant effect.
\item \textit{Annular clearance:} Inversely affects the amplitude of the collar signature.
\item \textit{CCL coil turns:} Directly proportional to the amplitude of the collar signature.
\item \textit{CCL coil length:} Alters the morphology of the collar signature.
\item \textit{CCL magnet size:} Exhibits no significant effect.
\item \textit{RIH speed:} Directly proportional to the amplitude of the collar signature, but inversely proportional to its temporal width.
\end{itemize}

Under varying well conditions, the proposed method remains applicable by adjusting the gain of the PGA, as the fundamental characteristics of the collar signatures remain consistent across different wells. However, the algorithmic parameters must be calibrated to accommodate different CCL coil lengths and RIH speeds, as these operational factors directly influence the morphological characteristics of the signature.

\subsection{Application Scope}

Governed by the fundamental principles of collar signature generation, the proposed DTPPMP system is strictly applicable to cased-hole oil and gas wells equipped with ferromagnetic casing strings. Because the method relies solely on these material properties rather than gravity or orientation, it supports operations across diverse well types, including vertical, deviated, and horizontal wells.
Furthermore, relying on the principles of depth correlation, an accurate casing tally for the target well is indispensable. Additionally, while the proposed method is specifically designed for compatibility with both wireline and wireless perforating systems, it can be adapted for other tethered or untethered conveyance methods by adjusting the algorithmic parameters.
Finally, the system requires an RIH speed of approximately \SIrange{6}{8}{\kilo\meter\per\hour}, as the dynamic thresholding is optimized for this specific range. Significantly higher speeds may necessitate further calibration of the algorithm parameters.

The proposed method eliminates the reliance on surface equipment, thereby reducing both equipment footprint and personnel requirements. This characteristic is particularly advantageous for integration into self-destructing, disposable perforating tools. Moreover, the proposed approach contributes to enhancing the automation and intelligent execution of perforating operations, ultimately reducing operational costs, safety risks, and potential environmental impacts.

Notably, the proposed physical plausibility verification algorithm is orthogonal to specific collar recognition techniques. Therefore, it can be integrated with various collar-based depth measurement methods to further reduce false positive and false negative rates, ultimately improving the overall accuracy of collar recognition and depth measurement.

\subsection{Limitations and Future Work}

The availability of historical and field test data is currently limited due to the logistical challenges of securing access to operational wells. Additionally, the casing collar recognition and physical plausibility verification algorithms present opportunities for further optimization, which could ultimately enhance the overall performance of the system.

Future research should focus on conducting additional field tests and expanding the dataset to further refine the algorithmic performance. Moreover, improving the speed adaptability of the system and implementing adaptive parameter adjustment mechanisms represent key priorities for subsequent development. Furthermore, the integration of tiny machine learning (TinyML) techniques offers a promising approach to optimizing both casing collar recognition and physical plausibility verification, thereby increasing depth measurement accuracy. Finally, as autonomous perforating technologies continue to evolve, the incorporation of sensor fusion and multi-sensor arrays constitutes a highly valuable direction for future investigation.

\section{Conclusions}\label{sec:con}

This work proposes the DTPPMP system, a novel automatic self-locating perforating system powered by an ARM Cortex-M7-based microcontroller. The system enables real-time, in-situ, and precise depth measurements for perforating at designated perforating intervals. These capabilities are achieved by processing CCL signals and identifying collar signatures using a lightweight, dynamic-threshold-based collar recognition algorithm coupled with a physical plausibility verification algorithm. Within this framework, the physical plausibility verification algorithm is orthogonal to the collar recognition algorithm. This orthogonality allows it to be integrated with various collar recognition methods, yielding an expected improvement of approximately \SIrange{3}{5}{\percent} points in the F1-score of the collar recognition task.

Field tests have verified the system's capability to perform accurate depth measurements and execute perforations at designated intervals. The system successfully recognizes casing collar signatures with an F1-score of \SI{98.6}{\percent} at a throughput of \SI{1000}{\sample\per\second}. The recognition time requires only \SI{1.5}{\micro\second} per sample while maintaining robust performance, representing a practical step toward automatic self-locating perforating operations.

Ultimately, this work presents a methodology that contributes to the realization of automatic self-locating perforating systems. These developments represent a meaningful step toward the automation and intelligent control of perforating operations, with the potential to reduce reliance on surface equipment and mitigate associated operational costs and safety risks.



\bibliographystyle{IEEEtran}
\bibliography{reference}

\vfill
\newpage

\begin{IEEEbiography}[{\includegraphics[width=1in,height=1.25in,clip,keepaspectratio]{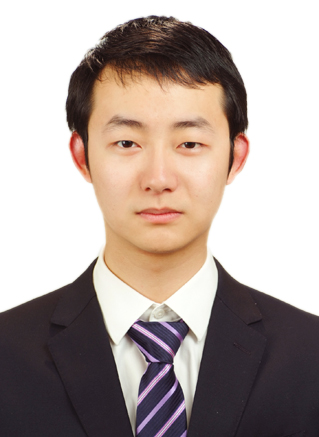}}]
~
Si-Yu Xiao received the B.S. degree in microelectronic science and engineering with the University of Electronic Science and Technology of China (UESTC), Chengdu, China, in 2018. He is currently pursuing the Ph.D. degree in electronic science and technology at UESTC.
His research interests include digital circuit design, edge computing, artificial intelligence, and their applications.
\end{IEEEbiography}

\begin{IEEEbiography}
[{\includegraphics[width=1in,height=1.25in,clip,keepaspectratio]{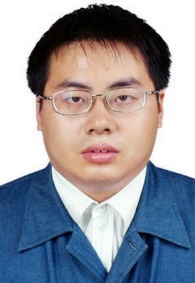}}]
~
Guo-Hui Ren is currently a Professor-Level Senior Engineer at Southwest Branch of China National Petroleum Corporation Logging Company Ltd., Chongqing, China.
His current research includes perforation optimization design software, new perforation technologies, and research on new methods.
\end{IEEEbiography}

\begin{IEEEbiography}[{\includegraphics[width=1in,height=1.25in,clip,keepaspectratio]{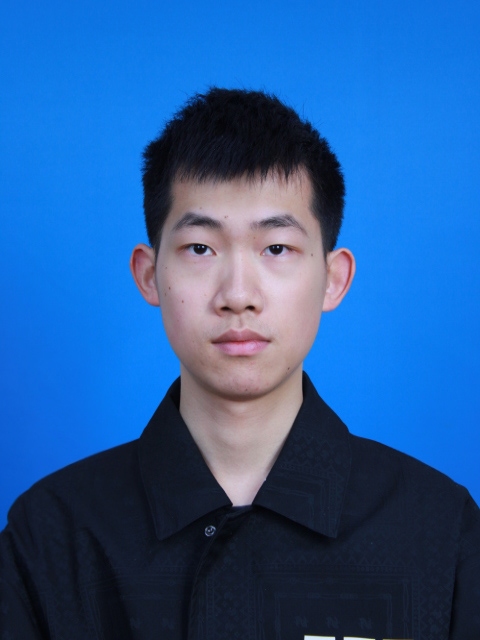}}]
~
Tian-Hao Mao received the B.S. degree in integrated circuit design and integrated systems from the University of Electronic Science and Technology of China (UESTC), Chengdu, China, in 2025. He is currently pursuing the Ph.D. degree in integrated circuit science and engineering at UESTC.
His academic interests focus on digital circuit design, system-on-chip (SoC) architecture, and algorithm optimization for hardware acceleration.
\end{IEEEbiography}

\begin{IEEEbiography}[{\includegraphics[width=1in,height=1.25in,clip,keepaspectratio]{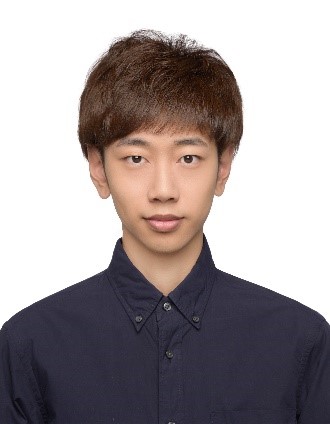}}]
~
Yu-Qiao Chen received the B.S. degree in microelectronic science and engineering from the University of Electronic Science and Technology of China, Chengdu, China, in 2025, where he is currently pursuing the M.S. degree in integrated circuit science and engineering.
His research interests cover digital circuit design, system-on-chip (SoC), and neural network algorithms.
\end{IEEEbiography}

\begin{IEEEbiography}[{\includegraphics[width=1in,height=1.25in,clip,keepaspectratio]{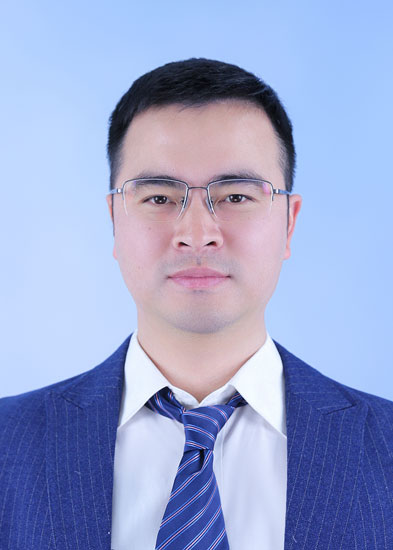}}]
~
Yi-An Liu (Member, IEEE) received the Ph.D. degree in microelectronics from the University of Electronic Science and Technology of China, Chengdu, China, in 2022.
His research interests focus on neural devices and networks.
\end{IEEEbiography}

\begin{IEEEbiography}[{\includegraphics[width=1in,height=1.25in,clip,keepaspectratio]{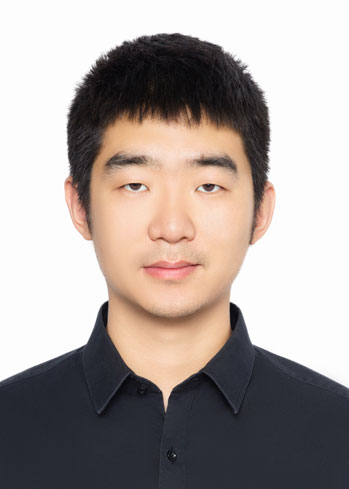}}]
~
Jun-Jie Wang received the Ph.D. degree in microelectronics from the University of Electronic Science and Technology of China (UESTC), Chengdu, China, in 2021.
He is currently a Researcher at the UESTC. His current research interests include digital circuit design, nonvolatile memory devices, and their applications in artificial intelligence.
\end{IEEEbiography}

\begin{IEEEbiography}[{\includegraphics[width=1in,height=1.25in,clip,keepaspectratio]{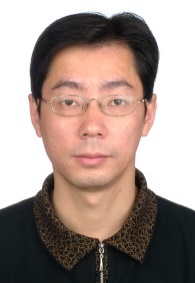}}]
~
Kai Tang is currently a Professor-Level Senior Engineer at Southwest Branch of China National Petroleum Corporation Logging Company Ltd., Chongqing, China.
His current research includes perforation engineering software development, new perforation technologies, and methods.
\end{IEEEbiography}

\begin{IEEEbiography}
[{\includegraphics[width=1in,height=1.25in,clip,keepaspectratio]{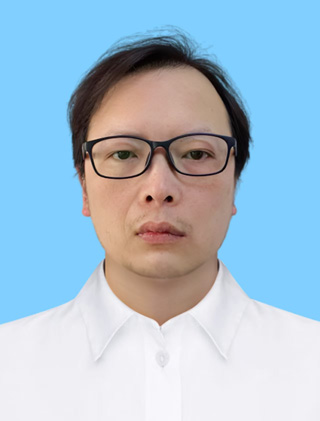}}]
~
Xin-Di Zhao received the Bachelor of Engineering degree in software engineering from Southwest Petroleum University, Chengdu, China, in 2010.
He is currently a Level 2 Engineer of perforation technology research at Southwest Branch of China National Petroleum Corporation Logging Company Ltd., Chongqing, China. His current research includes new perforation tools, technologies, and methods.
\end{IEEEbiography}

\begin{IEEEbiography}
[{\includegraphics[width=1in,height=1.25in,clip,keepaspectratio]{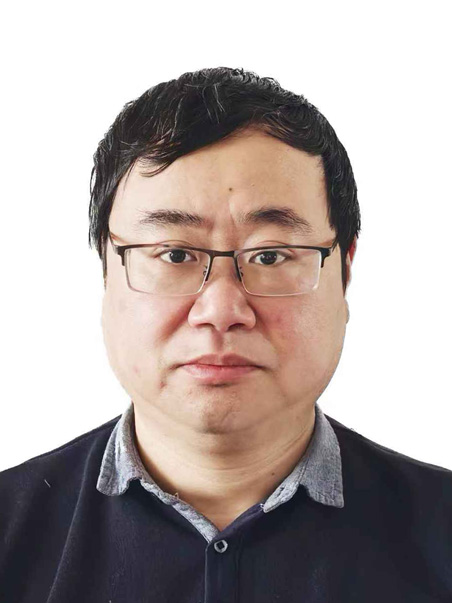}}]
~
Zhi-Jian Yu received the B.S. degree from the University of Electronic Science and Technology of China (UESTC), Chengdu, China, in 2000.
He is currently with Chengdu Original Dynamics Technology Company Ltd., Chengdu, China. His current research interests include applied electronics.
\end{IEEEbiography}

\begin{IEEEbiography}
[{\includegraphics[width=1in,height=1.25in,clip,keepaspectratio]{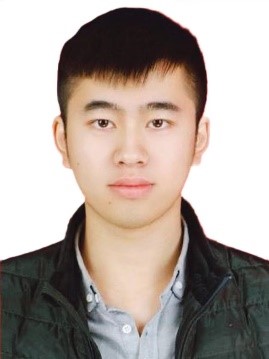}}]
~
Shuang Liu received the Ph.D. degree in microelectronics from the University of Electronic Science and Technology of China, Chengdu, China, in 2023.
He is currently a Post-Doctoral Researcher at the University of Electronic Science and Technology of China. His current research interests include processing{-}in{-}memory circuits and neuromorphic systems.
\end{IEEEbiography}

\begin{IEEEbiography}
[{\includegraphics[width=1in,height=1.25in,clip,keepaspectratio]{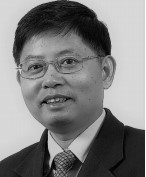}}]
~
Tu-Pei Chen is currently a tenured Faculty Member at the School of Electrical and Electronic Engineering, Nanyang Technological University, Singapore, where he has been teaching and conducting research in the field of microelectronics for over 25 years. His current research interests include on-chip ESD and latch-up protection, memory devices, memory-based computing (in-memory computing, neuromorphic computing), and thin-film transistors and applications.
\end{IEEEbiography}

\begin{IEEEbiography}
[{\includegraphics[width=1in,height=1.25in,clip,keepaspectratio]{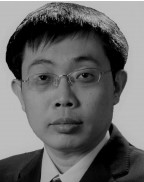}}]
~
Yang Liu received the B.Sc. degree in microelectronics from Jilin University, Changchun, China, in 1998, and the Ph.D. degree from Nanyang Technological University, Singapore, in 2005.
From May 2005 to July 2006, he was a Research Fellow with Nanyang Technological University. In 2006, he was awarded the prestigious Singapore Millennium Foundation Fellowship. In 2008, he joined the School of Microelectronics, University of Electronic Science and Technology of China, Chengdu, China, as a Full Professor. He is the author or co-author of over 130 peer-reviewed journal articles and more than 100 conference papers. He has been awarded one U.S. patent and more than 30 Chinese patents. His current research includes memristor neural network systems, neuromorphic computing ICs, and AI-RFICs.
\end{IEEEbiography}

\vfill

\end{document}